\documentclass[12pt]{article}

\usepackage{amsmath,amsthm}
\usepackage{color,soul}
\usepackage{natbib}
\usepackage{hyperref}
\usepackage{booktabs}
\usepackage{amsfonts}
\usepackage{graphicx}
\usepackage[export]{adjustbox}
\usepackage{cleveref}
\usepackage{pdflscape}
\usepackage{multirow}
\usepackage[small,bf]{caption}
\usepackage{subcaption}
\usepackage{setspace}
\usepackage[stable]{footmisc}
\usepackage{bm}
 
\usepackage{etoolbox}
\usepackage{needspace}
\usepackage{amsmath}
\usepackage{amsthm}
\usepackage{hyperref}
\usepackage{amsfonts}
\usepackage[a4paper]{geometry}
\newgeometry{left=3.5cm, right=2.5cm}
\usepackage{graphicx, multirow}
\usepackage{caption}
\usepackage{subcaption}

\usepackage{booktabs}

\title{Cyclical properties of supply-side and demand-side shocks in oil-based commodity markets\thanks{For estimation of the frequency dependent connectedness measures introduced by this paper, we provide the package \texttt{frequencyConnectedness} in \textsf{R} software. The package is available at \url{https://github.com/tomaskrehlik/frequencyConnectedness}. Support from the Czech Science Foundation (GA\v CR) under the 16-14179S grant is gratefully acknowledged.}}

\author{%
Tom\'{a}\v{s} {\sc K\v{r}ehl\'ik}$^{\rm a,b}$\thanks{Corresponding author, Tel. +420(724)091926, Email address: tomas.krehlik@gmail.com}, and
Jozef {\sc Barun\'{i}k}$^{\rm a,b}$
\vspace{5mm} \\
\small $^{\rm a}$ Institute of Economic Studies, Charles University, \vspace{-0.5mm}\\  \
\small Opletalova 26, 110 00, Prague, Czech Republic \vspace{3mm} \\
\small $^{\rm b}$ Department of Econometrics, IITA, The Czech Academy of Sciences, \vspace{-0.5mm}\\
\small Pod Vodarenskou Vezi 4, 182 00, Prague, Czech Republic}

\begin{document}

\maketitle

\begin{abstract}

Oil markets profoundly influence world economies through determination of prices of energy and transports. Using novel methodology devised in frequency domain, we study the information transmission mechanisms in oil-based commodity markets. Taking crude oil as a supply-side benchmark and heating oil and gasoline as demand-side benchmarks, we document new stylized facts about cyclical properties of the transmission mechanism generated by volatility shocks with heterogeneous frequency responses. Our first key finding is that shocks to volatility with response shorter than one week are increasingly important to the transmission mechanism over the studied period. Second, demand-side shocks to volatility are becoming increasingly important in creating short-run connectedness. Third, the supply-side shocks to volatility resonating in both the long run and short run are important sources of connectedness.

\end{abstract}

\section{Introduction} 

Oil-based commodity markets are subject to continuous evolution because of permanent inflow of new technologies, ecological pressures, and geopolitical importance of the control of oil supplies. More importantly, oil-based commodities are of paramount importance to economic prosperity in both developed and developing countries because they constitute the most widely used source of energy; for illustration, in 2014, about 30\% of US energy consumption used petroleum-based fuels, of which about 70\% was used in transportation and the rest in industrial usage.\footnote{Lawrence Livermore National Laboratory, Energy Chart 2014: \url{https://flowcharts.llnl.gov/content/assets/docs/2014_United-States_Energy.pdf}.}

In this paper, we study cyclical properties of shocks to volatility propagating through petroleum markets. Focusing on the importance of modelling both overall (aggregate) and cyclical (disaggregate) risk relations, we document that overall risk is highly dynamic and that the connectedness measure provides an accurate way of assessing it. Decomposing the risk into frequency domains, we investigate the roles that various types of information with heterogeneous frequency responses play in creation of such risk. Specifically, macroeconomic announcements constitute prime examples with monthly frequency, while quarterly company results might have a long-run effect. On the contrary, a weather forecast will impact the system in the shorter run. In turn, these shocks will propagate through the market with different frequency responses. Frequency domain-based measures allow us to identify the importance of various types of such shocks in the creation of risk in the system. Our results demonstrate new stylized facts about cyclical properties of connectedness in petroleum-based products.

As a workhorse, we use a small vector autoregressive (VAR) system of realized volatilities with three commodities: crude oil (CO), heating oil (HO), and gasoline (HUXB). Crude oil is the raw material that is used to produce the heating oil, gasoline, and other petroleum-based products. This is important in part of our interpretation because shocks to the volatility of crude oil can help us identify the supply-side shocks. As our framework is relatively simplistic compared to the vastly specified VAR models that are standard in the literature of oil-commodity modelling, we discuss potential caveats in interpretation by relating our results to relevant literature. 

Our paper contributes to the literature by investigating the volatility transmission mechanism, with an accent on cyclical properties of the transmission generated by shocks with heterogeneous frequency responses. Our results hence shed light on long-run and short-run patterns that emerge. Our key contribution is twofold. First, we document several new stylized facts about information transmission effects: \emph{a)} we show the growing importance of information transmission effects of up to one week and the overall diminishing importance of the longer movements; \emph{b)} the relative contributions of supply-side and demand-side shocks are more pronounced in the long run than in the short run; \emph{c)} various geopolitical and economic events had different effects on short- or long-run information transmission mechanisms. Second, from the methodological point of view, we define and apply complementary directional connectedness measures to the previous work of \citet{barunik2015measuring}.  

The rest of the paper is structured as follows. First, we review the relevant literature in the following section. Then, we continue with the explanation of our methodology, continuing next to the data, interpretation strategy, empirical results, and their discussion.

\section{Literature review}

From a theoretical point of view, we employ the framework of \citet{ross1989information} who identifies the standard deviation of price to be synonymous to the rate of information flow inside the standard martingale-based price models. Hence, modelling the connection between volatilities is synonymous with modelling connections between information flows, sometimes called the information transmission mechanism. We take this methodology a step further and impose generally reasonable assumptions on our system, inducing the information transmission mechanism to become synonymous with the systemic risk. Such a model provides us with a practical assessment of how risks depend upon each other.

The observation that volatility plays an important proxy for the information mechanism and systemic risk has been recently applied in conjunction with the general connectedness/network methodology of \citet{diebold2012better,diebold2009,diebold2013measuring} to assess how information transmissions work in various cases \citep{kocenda2015volatility,zhou2012volatility,alter2014dynamics,awartani2013dynamic}. An important extension of this framework by \citet{barunik2015measuring} studies the frequency properties of generalized impulse response functions, providing a complementary picture of the structure of the system. Specifically, \citet{barunik2015measuring} argue that frequency dynamics is insightful for studying connectedness, as shocks with heterogeneous frequency responses create linkages with various degrees of persistence. Economically, this analysis allows us to study whether most of the future volatility will happen in short-run movements or in one continuous long-run move. Additionally, the methodology allows us to evaluate what type of shocks are the most important for the risk of the system. In this paper, we use the framework within a simple three variable VAR that is fit locally to a system of volatilities following the assumption of local stationarity as in \citep{stuaricua2005nonstationarities}. Contributing to the methodology of measuring information transmission mechanisms, we define the directional connectedness measures within this frequency framework.

The previous literature has been greatly interested in oil commodity markets, perhaps because they play a prominent role and hence are an important part of the US economy. Numerous studies investigate the relationship between business cycles and the price of crude oil. The beginning of this literature dates back to the work of \citet{hamilton1983oil}, who concentrated on an interplay between the price surges of crude oil and macroeconomic crises in the US. Since then, multiple authors have studied similar relations. \citet{hamilton1996happened} revisited the macroeconomic relation that became quite unstable after the year 1986. A subsequent work of \citet{hamilton2009causes} compares and contrasts the oil shock of 2007-2008 and concludes that in comparison with the previous oil shocks, this shock was caused by strong demand meeting stagnating production.

Other aspects of the macroeconomic and oil relationship were prominently investigated in the works of \citeauthor{kilian2010explaining}. \citet{kilian2009not} suggests a decomposition of shocks affecting oil into three distinct shocks: crude oil supply shocks, shocks to the global demand for all industrial commodities, and demand shocks that are specific to the global crude oil market. The author innovatively uses freight cargo fares to benchmark the global economic activity and subsequently uses this variable to clean the oil prices from the global economic activity. \citet{kilian2014role} investigate the role of inventories and speculative trading in crude oil. They refute claims that the 2003-2008 surge in prices was caused mainly by speculations, proposing instead that it was caused by the unexpected increase in world oil consumption. Moreover, they claim that the short-run price elasticity of oil demand is much higher than traditional estimates from dynamic models would suggest because the models do not account for the endogeneity of the price of oil.

\citet{kilian2010explaining} studies interaction of the crude oil market with the US retail gasoline market using five variables to structurally identify all shocks. The variables are as follows: price of crude oil, price of gasoline, global oil production, global real economic activity, and US consumption of gasoline. Carrying out thorough impulse response analysis, \citet{kilian2010explaining} answers several questions: what is the structure of demand and supply shocks since March of 1974, how do the prices respond to demand and supply shocks, how does the consumption respond to the shocks, and how have price fluctuations since 2002 happened? The approach is structurally more elaborate; in particular, it disentangles a higher number of shocks that we cannot underpin in our approach. The study shows that in the short run, most of the price movements are caused by refining shocks. However, in the long run, fluctuations are driven by demand shocks and shocks to the business cycle. The refining shocks play only a very small role in the long run. Regarding the consumption of crude oil, most of it is driven by demand shocks.

\citet{cashin2014differential} try to identify supply and demand shocks to the oil price within a global VAR (GVAR) model that is estimated for 38 countries. They include more countries inside the GVAR than was previously done. However, they concentrate more on macroeconomic effects than on the relations between the two commodities.

Another important strand of literature investigating petroleum concentrates on the question of price elasticity. This issue is especially important because in recent literature, the elasticities tie the increases in volatility to changes in prices. If the price elasticity is relatively low, large movements of prices are needed to clear the market. One of the most recent attempts \citet{hughes2006evidence} evaluated the short-run price elasticities of gasoline demand. The derived short-run gasoline demand elasticities are very close to zero in a sample similar to ours. \citet{guntner2014oil} concentrates on demand-driven price changes in the time span 1975--2011. The authors derive consistent short-run country-specific price elasticity and conclude that the supply elasticities seem to be indistinguishable from zero. Most relevant for our paper, \citet{baumeister2013role} investigate the reason behind the increase in volatility since the second half of the 1980s. They show that the likely explanation is that the price elasticities are very low and that both demand and supply shocks have declined over time. On the one hand, they find that since the invasion of Kuwait in 1990, exogenous supply shocks have declined steadily. On the other hand, they confirm the finding from \citet{kilian2014role} that the demand shocks were most probably the important force behind price fluctuations during the period 1974-2009. They use Bayesian TVP-VAR to explore their hypothesis.

An empirically interesting evolution of prices happened during the period of 2007-08 when the prices of oil spiked. The proponents of the theory that speculative trading caused the evolution argue that the trading strategy is to buy in on near-term future contracts and sell before expiry. If the prices trend upwards, the proceeds can be invested in another round of trading without ever touching the commodity itself. If more and more investors seek this strategy, the artificial demand will drive up the prices of the futures, inducing a speculative bubble in the oil price. \citet{singleton2013investor,masters2008accidental} are the biggest proponents of the tangible involvement of hedge fund investments in the 2007-08 boom bust in oil prices.

There is, however, little literature that investigates the volatility connectedness effects of oil-related markets. \citet{li2016exogenous} investigate the (mean and volatility) information transmission using the EGARCH model within VECM specification. They include exogenous shocks such as the S\&P 500, VIX, gold price, TED spread, and US dollar. They make two important conclusions. First, they find evidence that there is an important volatility transmission mechanism that is moreover quite different before and after the crisis of 2008. Second, they show that exogenous variables can have important effects on the volatility transmission mechanism. \citet{kocenda2015volatility} study the same data as we do with similar methodology; however, they concentrate on uncovering the asymmetric volatility connectedness effects. They find an asymmetric effect in the information transmission mechanism that is dynamic over time. The asymmetry measure is significantly higher during the pre-2008 crisis period than afterwards. Lastly, \citet{Maghyereh_2016} present a study of connectedness between oil and equity markets. Using implied volatility as a proxy for the latent volatility process, they find that the flow of connectedness from the oil to the equity market strongly dominates the other direction.

\section{Methodology}

There are multiple reasons one should believe that connectedness in volatility systems and more generally in financial and commodity markets should be different at different levels of persistence. The general reason is that agents on financial markets are not all alike; some have preferences in longer horizons, and some have short-term preferences. This diversity of utility functions is necessarily lost in aggregate measures that ignore it. Additionally, information might have various frequencies by itself. As argued earlier, there are quarterly reports about financial earnings and yearly reports concerning the whole economy. Much can be gained in terms of structural understanding of economic models when proper spectral tools are used to construct models that can discern between different horizons. 

The economic literature has recently recognized this shortcoming and started to address it in multiple ways. \citet{Dew_Becker_2016} use spectral methods to show the implication of usage of various utility functions within asset pricing. \citet{bandi2015business} uses local spectral methods to investigate the traditional finding of \citet{bansal2004risks} that long-run returns can be predicted better than short-run returns. The methodology that follows is our contribution to that literature; we investigate the spectral patterns within the information transmission mechanisms and, in our particular case, systemic risk.

\subsection{Cyclical properties of shock responses}

The measure of connectedness, much like any model-based measure, necessitates an assumption about the data-generating process. \citet{diebold2012better} use probably the most general and versatile assumption---the vector-autoregressive (VAR) model. Hence, let us have vector $X_t$ that holds volatilities of $k$ assets at time $t$ and assume the dynamics of $X_t$ follow

\begin{equation}
	X_t = \Phi(L) X_t + \mathbf{\epsilon}_t,
\end{equation}
where $\Phi(L)$ is a lag polynomial generating stable VAR system, and $\mathbf{\epsilon}_t \sim N(0, \Sigma)$. The coefficients of this model can be estimated equation-by-equation using ordinary least squares, which also corresponds to the maximum likelihood estimate.

The stationary system can be rewritten in a moving average (MA)-representation as

\begin{equation}
	X_t = \Psi(L)\mathbf{\epsilon}_t= \sum_{i=1}^{\infty} \Psi_{i} \mathbf{\epsilon}_{t-i} + \mathbf{\epsilon}_t.
\end{equation}

Based on these estimates, using the generalized VAR identification scheme of \citet{pesaran1998generalized}, we can compute the generalized impulse responses to shock in variable $j$ at time $t+h$ as
\begin{equation}
	\text{GIRF}_j (h) = \sqrt{\Sigma_{j,j}} \Psi_h \Sigma e_j,
\end{equation}
where $e_j$ is a $k$-length vector with 1 at a position $j$ and 0 otherwise, $\Psi_h$ denotes the corresponding coefficients of Wold decomposition at the lag $h$, and $\Sigma_{j,j}$ is a $j$th diagonal element of $\Sigma$ matrix. This generalized impulse response can be further leveraged to construct a generalized forecast error variance decomposition (GFEVD) given by
\begin{equation}
	\left( \theta_H \right)_{i,j} = \frac{\Sigma_{j,j}^{-1}\sum_{h = 0}^{H}(\Psi_h \Sigma)^2_{i,j}}{\sum_{h=0}^{H} (\Psi_h \Sigma_{\epsilon} \Psi_h')_{i,i}},
\end{equation}
where $H$ defines the horizon, i.e., how many periods ahead we are cumulating. The relation gives the shares of forecast error variances in variable $i$ due to shock to variable $j$. 

Inspired by similar approaches in the literature, \citep{stiassny1996spectral,Dew_Becker_2016}, \citet{barunik2015measuring} use spectral methods to further investigate the implied unconditional connectedness relations in the frequency domain. The decomposition is achieved by an observation that the spectral behavior of series $X_t$ can be described by its frequency response function
\begin{equation}
    S_X(\omega) = \sum_{h=0}^{\infty} E(X_t X_{t-h}) e^{-i h \omega} = \Psi(e^{-i h \omega}) \Sigma \Psi(e^{i h \omega}),
\end{equation}
where $\Psi(e^{-i h \omega}) = \sum_{h=0}^{\infty} \Psi_h e^{-i h \omega}$. These frequency response functions can be used to decompose the generalized impulse response functions.

Based on these observations, the authors derive the GFEVD on frequency $\omega$ as
\begin{equation}
    \left( \theta (\omega) \right)_{i,j} = \frac{\Sigma_{j,j}^{-1} \sum_{h = 0}^{\infty}(\Psi(e^{-i h \omega}) \Sigma)^2_{i,j}}{\sum_{h=0}^{\infty} (\Psi(e^{-i h \omega}) \Sigma \Psi(e^{i h \omega}))_{i,i}}.
\end{equation}
Note also that the horizon $H$ does not play an important role, as we work with unconditional GFEVD, taking infinite horizon relations. In a discrete setting, this is mimicked by taking sufficiently large $H$.

Standardizing as
\begin{equation}
    \left( \widetilde{\theta}(\omega) \right)_{i,j} = \left( \theta(\omega) \right)_{i,j} / \sum_{j=1}^{k} \left( \theta(\omega) \right)_{i,j},
\end{equation} \citet{barunik2015measuring} arrive at a connectedness table at chosen frequency $\omega$. The table provides us very condensed information about the system in the sense that it measures how the shock to variable $j$ influences variable $i$.

Because the connectedness tables at individual frequencies $\omega$ are both un-informative and infeasible in a discrete setting, \citet{barunik2015measuring} propose to accumulate tables over frequencies $\omega$ in such a manner that connectedness tables are formed at informative frequency bands, e.g., all frequencies that correspond to movements shorter than one week are accumulated into one connectedness table. For that purpose, let us define the connectedness table on an arbitrary frequency band $d = (a, b)$ as

\begin{equation}
    \left( \widetilde{\theta}_{d} \right)_{i,j} = \int_{a}^{b} \left( \widetilde{\theta}(\omega) \right)_{i,j} d \omega.
    \label{eq:fevdond}
\end{equation}
This entity allows us to proceed to the definition of connectedness measures. 

\subsubsection{Connectedness measures in the frequency domain}

Inspired by the measures defined by \citet{diebold2012better} on the GFEVD, one has to contemplate carefully how to extend the definitions to the frequency domain. Straightforwardly, we can apply the measures on the connectedness tables corresponding to any arbitrary frequency band $d$ as described in Equation~\ref{eq:fevdond}. Hence, for the \textsc{overall} connectedness, we have \begin{equation}
    \mathcal{C}^d = \frac{\sum_{i=1,i\neq j}^k \left( \widetilde{\theta}_d \right)_{i,j}}{\sum_{i,j} \left( \widetilde{\theta}_d \right)_{i,j}} = 1 - \frac{\sum_{i=1}^k \left( \widetilde{\theta}_d \right)_{i,i}}{\sum_{i,j} \left( \widetilde{\theta}_d \right)_{i,j}},
\end{equation}
where $d$ denotes the respective frequency band. These connectedness measures, however, only pertain to the movements happening inside the spectral band, disregarding the aggregate nature of the series. For example, $\mathcal{C}^d = 0.9$ tells us that within the frequency band, there are strong connections without any relation to the aggregate connectedness measure, which can be relatively low. Therefore, we call these measures within measures as in \emph{within spectral band} measures.

Similarly, we can define for each asset $i$ a measure of variance contributed by other variables $i \neq j$ that can be constructed as 
\begin{equation}
    \mathcal{C}_{i \leftarrow \cdot}^d = \sum_{j=1,i\neq j}^k \left( \widetilde{\theta}_d \right)_{i,j},
\end{equation}
the so-called within \textsc{from} connectedness on the spectral band $d$, and contribution of asset $i$ to variances of other variables as
\begin{equation}
    \mathcal{C}_{i \to \cdot}^d = \sum_{j=1,i\neq j}^k \left( \widetilde{\theta}_d \right)_{j,i},
\end{equation}
the so-called within \textsc{to} connectedness on the spectral band $d$. These two measures show how other assets contribute to the risk (in case the modeled variables are variances) of asset $i$, and how the asset $i$ contributes to the riskiness of others on the frequency band $d$.

The third measure shows the discrepancy between how much of the variance is received and how much is imposed. This so-called within \textsc{net} connectedness is computed as
\begin{equation}
    \mathcal{C}_{i,\text{net}}^d = \mathcal{C}_{\cdot \to i}^d - \mathcal{C}_{i \leftarrow \cdot}^d.
\end{equation}

In our case, the measure can be interpreted easily as whether the asset induces more risk than it receives from the other elements of the system. These three measures concisely describe the behavior of the individual elements within the band $d$. 

Apart from overall characteristics, we might be interested in pairwise relations of risk that can further be described by the \textsc{pairwise} connectedness

\begin{equation}
    \mathcal{C}_{i,j}^d = \left( \widetilde{\theta}_d \right)_{j,i} - \left( \widetilde{\theta}_d \right)_{i,j}.
\end{equation}
We will leverage this measure to describe more thoroughly the relation of products and raw material in petroleum markets.

To reach a measure that shows us the contribution of the given frequency band $d$ to the aggregate measure, the within measures need to be weighted. For a better illustration, it is helpful to think about the following example. 

Let us have two systems that both have very strong within connections in the short term and no within connections in the long term. However, the aggregate behavior of system number one is characterized by long-term movements (as in an AR process with very high coefficients), and the aggregate behavior of system number two is characterized by short-term movements (as in an AR process with negative coefficients). Because the variance in the first system is created mostly by long-term movements that are unconnected, the system will not be (or only slightly) connected in the aggregate despite the strong within short-term connections. However, the second system will show strong connectedness because the connected short-term movements compose most of the behavior of the system.

This leads us to a straightforward extension of the measures. The aggregate measure on the frequency band $d$ is defined as

\begin{equation}
	\widetilde{\mathcal{C}}^d = \mathcal{C}^d \cdot \Gamma(d),
    \label{eq:reconstruction}
\end{equation}
where $\Gamma(d) = \sum_{i,j=1}^k \left( \widetilde{\theta}_{d} \right)_{i,j}/\sum_{i,j=1}^k \left( \widetilde{\theta} \right)_{i,j} = 1/k \sum_{i,j=1}^k \left( \widetilde{\theta}_{d} \right)_{i,j}$ is the contribution of frequency band $d$ to the overall behavior of the system, and $\mathcal{C}^d$ is the connectedness measure computed on the connectedness table $\widetilde{\theta}_d$. 

The frequency measures denoted with tildes have the property that if we sum them up over disjointed intervals that give a range of frequencies, the unconditional connectedness measure results, \emph{i.e.} $\sum_d \widetilde{\mathcal{C}}^d = \mathcal{C},$ where $\mathcal{C}$ is the total connectedness defined in \citep{diebold2012better}.

Both within-spectrum and overall measures are important in investigating relations within the system because they demonstrate change in the structure of the series versus change only where most of the movements are concentrated.

Apart from the measures alone, the spectral weight $\Gamma (d)$ can provide us with valuable information on how the within connectedness is transformed into the absolute frequency connectedness on a given frequency band $d$.

\section{Empirical results}

Further, we describe the evidence about the supply and demand shocks through the lens of the connectedness measures devised in the previous section. We restrict ourselves to presentation of the frequency decomposed results and their implications.

\subsection{Data and estimation procedure}

For the inquiry, we use futures prices of three commodities: crude oil (CO), heating oil (HO), and gasoline (HUXB). The gasoline futures contract is composed of two data series that are connected together because in 2006, the NYMEX changed the contracts for gasoline and substituted the unleaded gasoline (HU) contracts with the reformulated gasoline blendstock for oxygen blending (XB). For computation purposes, we use HU before 2006 and XB after 2006.

From the high-frequency irregularly spaced data\footnote{The data were obtained from Tick Data, Inc., which uses data from Globex.}, we extract 5-minute returns and compute a bi-power realized measure \citep{barndorff2004power} of volatility that moreover disentangles jumps from underlying volatility. We exclude trades executed on Saturdays and Sundays, U.S. federal holidays, December 24 to 26, and December 31 to January 2 because of low activity on these days that could lead to estimation bias. The sample spans September 1, 1987, to February 12, 2014. The Table~\ref{tab:desc_stats} reports summary statistics for the realized volatility, and Figure~\ref{fig:evolutions} depicts the logarithmic volatility used in the estimation procedure.

The computation of connectedness necessitates use of the underlying model. We use the standard VAR with two lags and a constant that is fit on logarithmic volatilities. The use of logarithms is preferred in this case, as we are looking for relationships that are modelled through coefficients within the VAR system, and the transformed series better underpins these relations. We experimented with other settings of the VAR model, such as various lags and inclusion of trend or constant terms, and found that the presented specification is robust for interpretation purposes. In our view, the presented model is the most parsimonious approximation of reality.\footnote{The results for other specifications are available from the authors upon request.} The model is fit to the series on a rolling window basis. Such a procedure approximates the statistical properties of locally stationary series as argued in \citet{stuaricua2005nonstationarities}.

We decompose the resulting connectedness measure into two frequency bands: $(\pi, \frac{\pi}{5})$, $(\frac{\pi}{5},0),$ that correspond to movements up to five days and movements of five days and longer. In our case, the latter is constrained by the length of the moving window, i.e., 500 observations, meaning two years.

In each of the figures presented later, we include grid lines that denote important geopolitical events that might have influenced the volatility and in many cases did. In chronological order, the events are Iraq invasion of Kuwait, Asian Crisis, Russian Flu, Terrorist attacks on 9/11, US invasion of Iraq, fall of Lehman Brothers, and Arab Spring.

\subsection{Possible sources of shocks in volatility}

As noted in the introduction, the singular shocks to volatility of the commodities are not synonymous with demand shocks in the case of gasoline or heating oil, or supply shocks in the case of crude oil. In this brief section, we investigate possible causes of volatility shocks and relate them to possible effects on the respective shocks within our system.

The oil-commodity market is subject to many influences. First, exogenous shocks to supply, such as wars in places of drilling sites, political pressures such as OPEC agreements, or revolutions in oil producing countries, will likely have a significant effect on prices of crude oil and hence its volatility. Similar shocks may influence the demand of oil products, mainly the bursts of economic bubbles, subsequent crises, or local or international regulations of fossil fuels. \citep{kilian2009not}

Second, elements that are endogenous to oil derivatives production and the nature of such production may cause changes in demand volatility. Namely, the amount of inventory will likely smooth out oil prices, as with sufficient inventories, refineries can wait until the short-term price changes disappear from the market, hence decreasing the short-term swings in volatility. Weather-related events might influence the functioning of refineries as well, as shown during the hurricane seasons in the US. \citep{kilian2014role}

Third, part of the literature suggests that there is an increased involvement of financial institutions within the oil commodity market. Supposing the involvement is tangible, the volatility of oil commodity prices will be affected any time the financial institutions start heavily rebalancing their portfolios. Thus, any economic shock will propagate to the system \citep{kilian2014role,singleton2013investor}.

Lastly, demand of oil products is dependent on how the world economy fares. Higher world growth means higher demand for energy fuels and in turn higher volatility of oil and its products. This commonality that influences both supply and demand, however, should only emerge in spectral weights of the series and not change the fundamental connections within the system, \emph{i.e.,} not change the systemic risk \citep{hamilton1983oil,hamilton1996happened,kilian2010explaining}.

In the subsequent results, we refer to the ensemble of shocks as either supply-side or demand-side shocks. We allow ourselves this comfort, as the purpose of the paper is not to uncover the precise sources of the shocks but to underpin the dynamic changes in the fundamental connections between the oil markets over time. The literature on the issue of identifying the precise sources of shocks has been presented in the literature review.

\subsection{Interpretation strategy}

In the interpretation that follows, one of the tacit assumptions is that positive shock to volatility in one of the assets in our systems cannot have a negative impact on the volatility of other assets, i.e., if there is a shock that increases the volatility of crude oil, the same shock cannot be a cause for decreased volatility in the other two assets.

This is important mainly because patiently observing the connectedness measure alone does not imply \emph{per se} that increased connectedness also means higher average volatility. On the contrary, if we had two assets connected in \emph{complementary} fashion, meaning that increases in one asset's volatility would be outweighed by decreases in the second asset's volatility, the average volatility in the system could be lower than for non-connected series. The measure evaluates the importance of cross coefficients in the VAR system whether negative or positive. To properly address this issue, we need theoretical assumptions that elicit meaning from the measure.

For this reason, let us look at how prices can be obtained for the individual assets in our case. Suppose that the prices of the three commodities are functions as follows
\begin{align*}
    p_{CO} =& f(\text{drilling cost}, \text{transportation cost}, \text{storage cost},\\ &\text{demand gasoline} + \text{demand heating oil} + \text{demand other}, \text{external shocks})\\
    p_{HO} =& f(p_{CO}, \text{refinement cost}, \text{transportation cost}, \text{storage cost}, \text{demand heating oil})\\
    p_{GO} =& f(p_{CO}, \text{refinement cost}, \text{transportation cost}, \text{storage cost}, \text{demand gasoline}).
\end{align*}
We are heavily inspired by the previous literature in the field that addressed the issues of the structure of the oil market.

Assuming the linearity of the price function, the derivation of the volatility of price is straightforward. Hence, the only way one might obtain lower volatility in one asset through shock to the other asset would be through covariance of elements of the pricing equation. However, we suggest that the increase in costs of one type never directly decreases the price of other components. This leads to non-negative covariances.

In this framework, a positive shock to the variance of one asset must have a positive impact. Therefore, we interpret the following results bearing in mind that increased connectedness in volatilities also means increased overall exposure to volatility and hence higher systemic risk.

\subsection{Overall frequency connectedness}
\label{subs:Overall connectedness}

\begin{figure}[thp]
  \center
  \includegraphics[scale= 0.5]{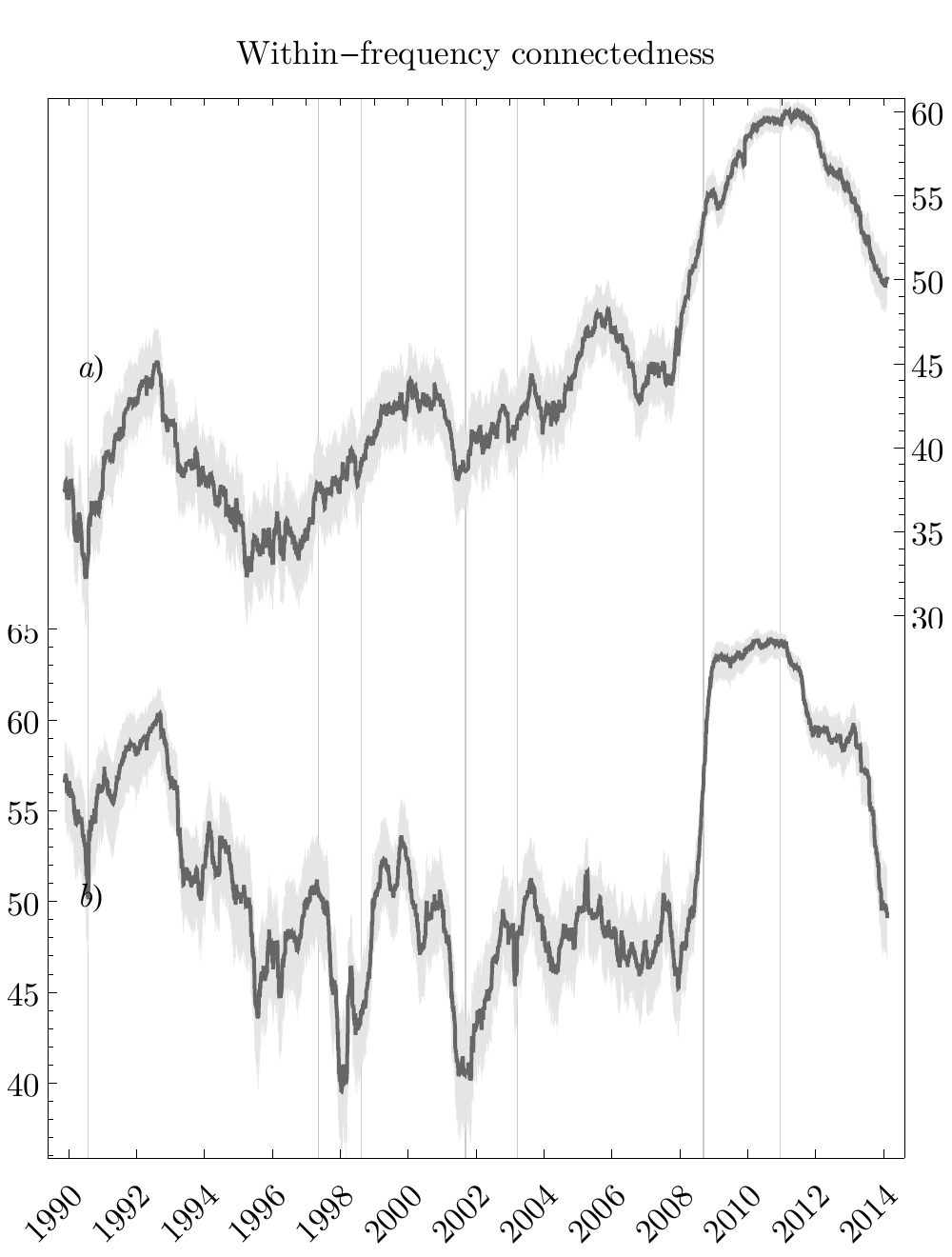}
  \includegraphics[scale= 0.5]{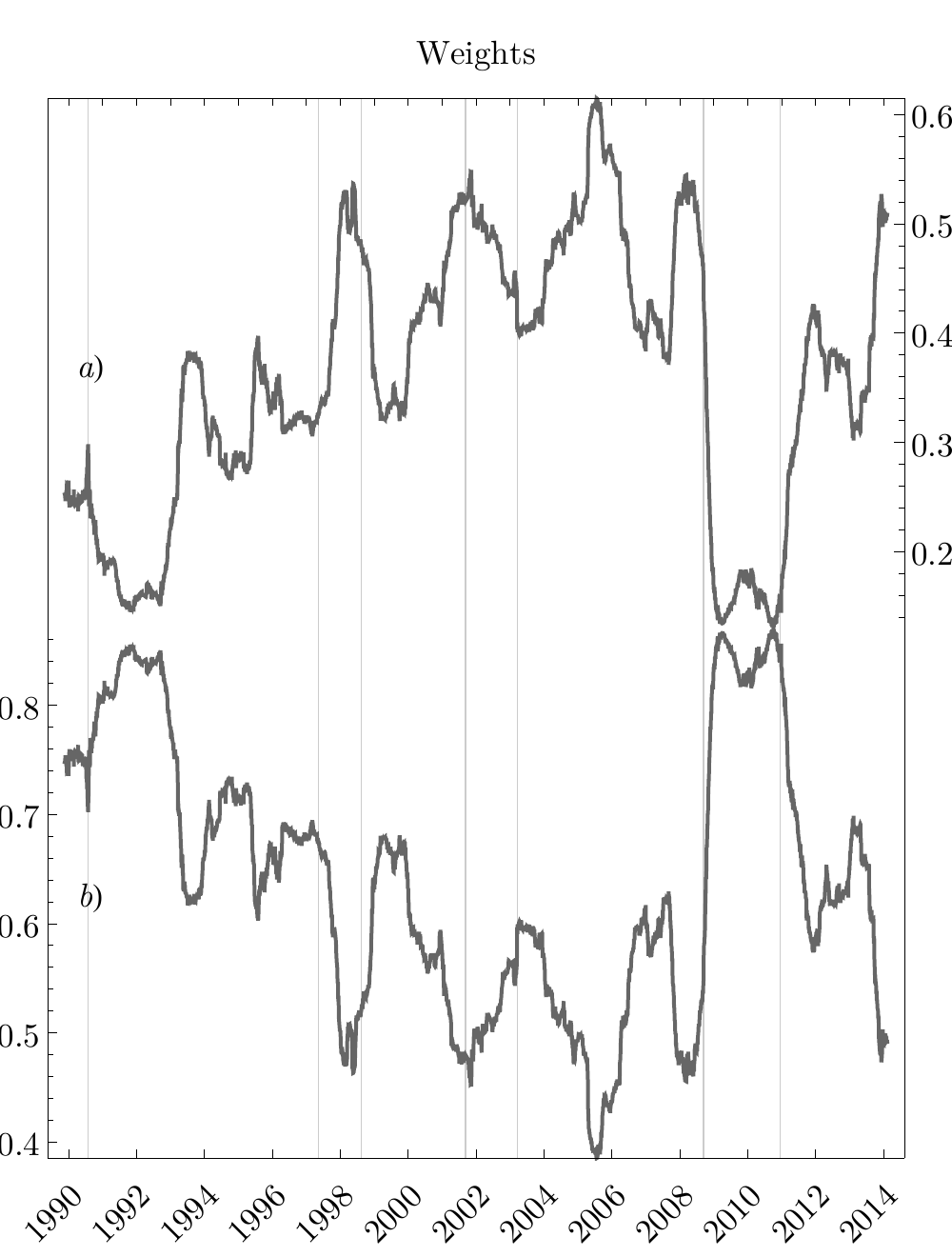}
  \includegraphics[scale= 0.5]{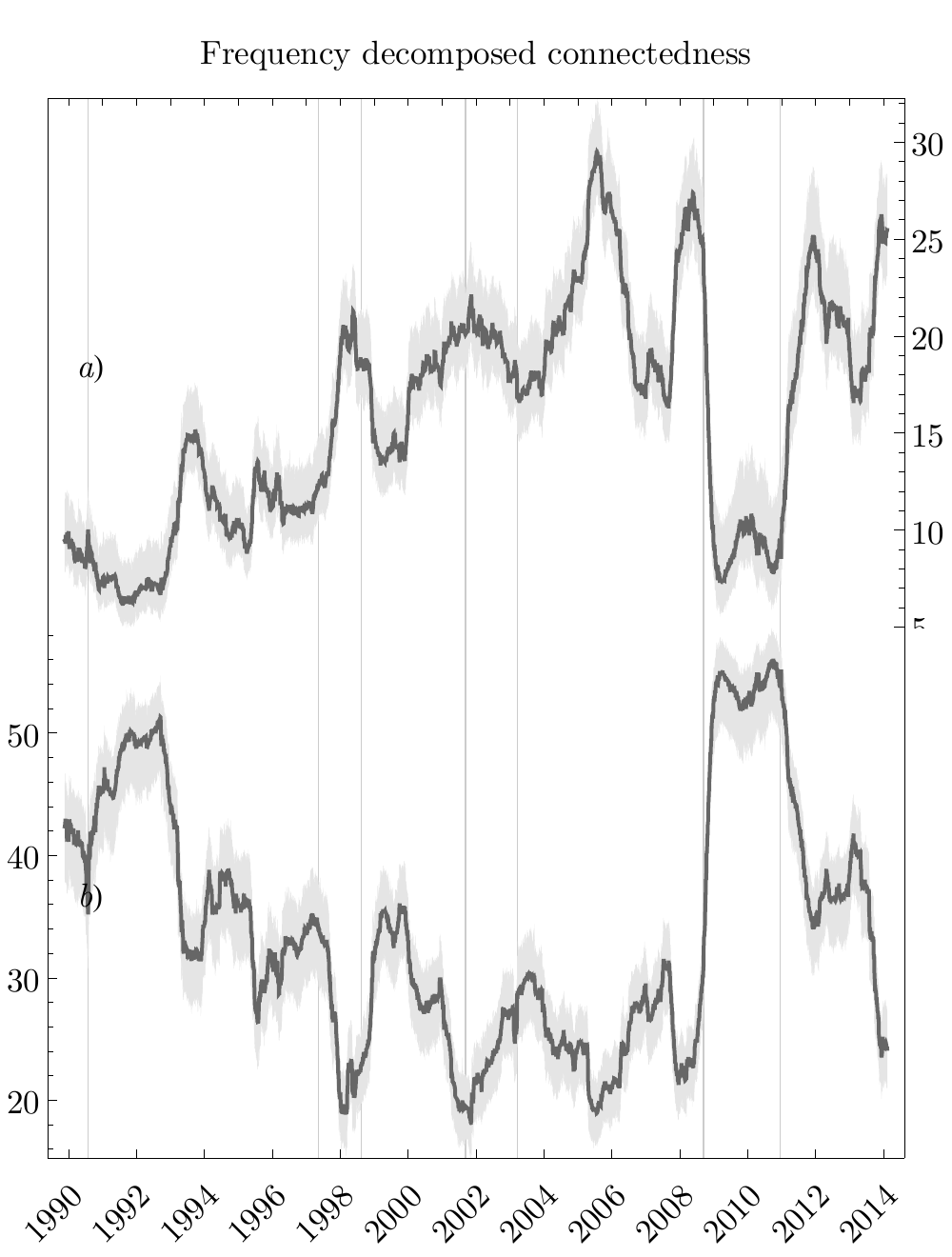}
\caption{Overall connectedness of crude oil, heating oil, and gasoline. The respective parts correspond to parts of Equation~\ref{eq:reconstruction}. The left figure shows the within connectedness measure ($C_d$), the middle figure shows spectral weights ($\Gamma(d)$), and the right figure shows the frequency connectedness ($\widetilde{C}_d$). The top lines denoted as \emph{a)} show the measures on the frequency band of up to a week (one to five days). The bottom lines denoted as \emph{b)} show the measure on the frequency band from one week to two years (six and more days to 500 days). The shaded 10\% confidence bands are based on parametric bootstrap.}
  \label{fig:overall}
\end{figure}

We start the empirical findings by the interpretation of the overall frequency decomposed connectedness. The rightmost picture in Figure~\ref{fig:overall} shows how the overall connectedness effect is decomposed into two parts. The short-term connectedness ranges from 10\% to 30\%, and in most, part of the sample is less important than the long-term connectedness that ranges from 20\% to 50\%. Before the crisis, a pattern emerges, where the short-run connectedness of the system is steadily increasing in importance, while the long-run connectedness decreases in importance. This, however, rapidly changes within the crisis period.

During the crisis period, the long-run connectedness surges, while the short-run connectedness decreases. It would be tempting to conclude that the fundamental connectedness within the system changes; however, looking at the other two pictures falsifies that hypothesis. Observing the within frequency band spillovers, we see that during the crisis, both long-term and short-term within connections actually increased. The weights decomposition then complements this picture, showing where such dynamics originate. It is not that the within connectedness would completely change during the crisis; it is the importance of the respective parts of the spectra that change. Because crisis periods are characterized by long slumps, the importance in the long run becomes prevailing during those periods.

Economically, the frequency decomposed connectedness is most important, as economic actors have to account for the nature of the series and the system connectedness as shown by the within-frequency connectedness. For qualitative insights into the behavior of the oil-products market and risk transmission within it, it is of paramount importance to look at the within-frequency connectedness, as it shows the fundamental risk transmission and the weights that show which part of the risk transmissions are important.

In the within connectedness, we can see then that the risk connectedness increases over time in the short run, i.e., holding positions that are shorter than one week are becoming more and more risky from the systemic risk point of view. Said otherwise, a singular increase in volatility in one of the elements in the series induces volatility in the other series in the short run more and more over time. However, disregarding crisis, there is a decreasing tendency in the amount of risk taken in the long positions over the long run. Moreover, as intuition would suggest, the systemic risk increases rapidly during the crisis; as uncertainty accumulates, any information is being scrutinized and processed more carefully, thus inducing increased connectedness.

\subsection{Directional frequency spillovers} 
\label{ssub:directional_frequency_spillovers}

\begin{figure}
  \center
  \begin{subfigure}[b]{0.49\textwidth}
        \includegraphics[scale= 0.35]{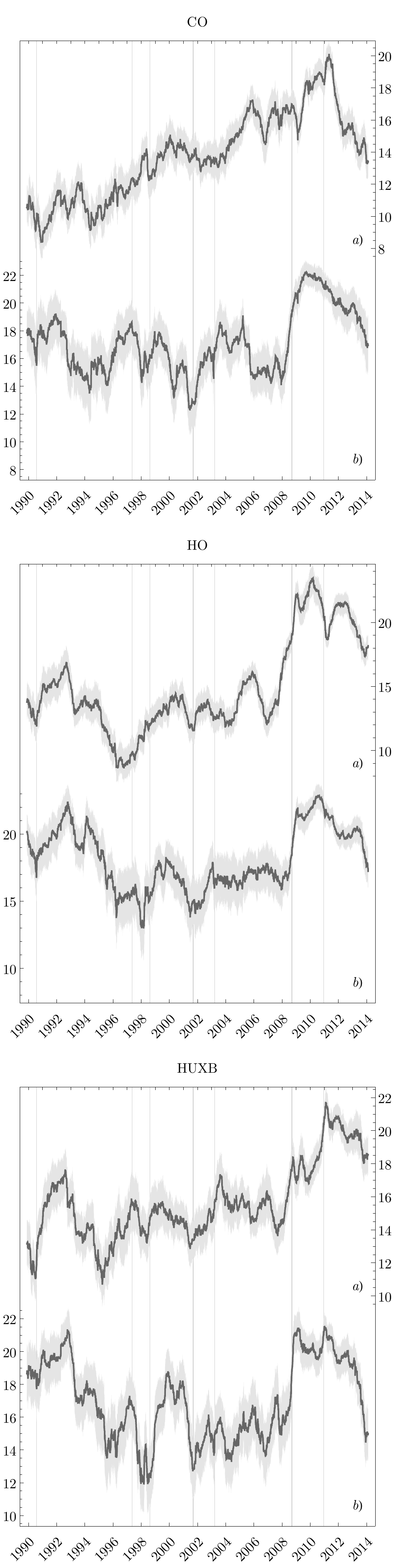}
        \includegraphics[scale= 0.35]{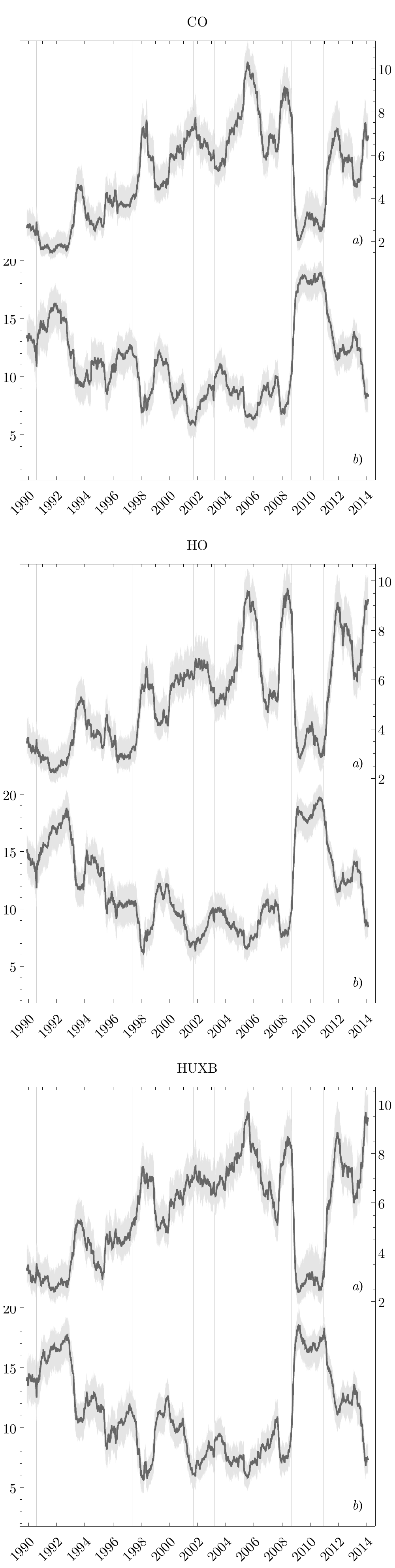}
        \caption{\textsc{From} connectedness}
        \label{fig:fromconnectedness}
  \end{subfigure}
  \begin{subfigure}[b]{0.49\textwidth}
        \includegraphics[scale= 0.35]{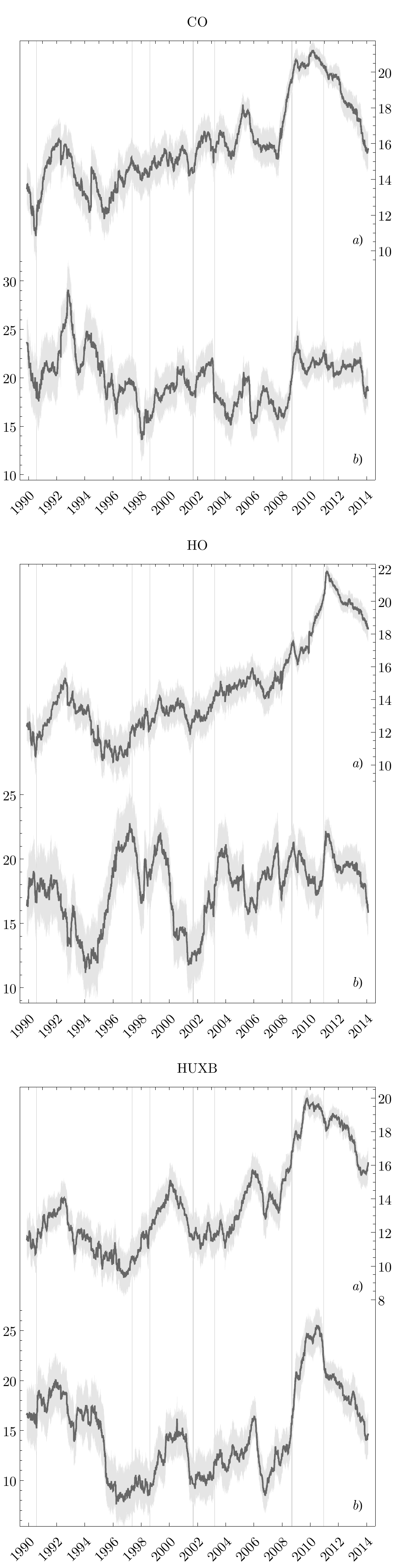}
        \includegraphics[scale= 0.35]{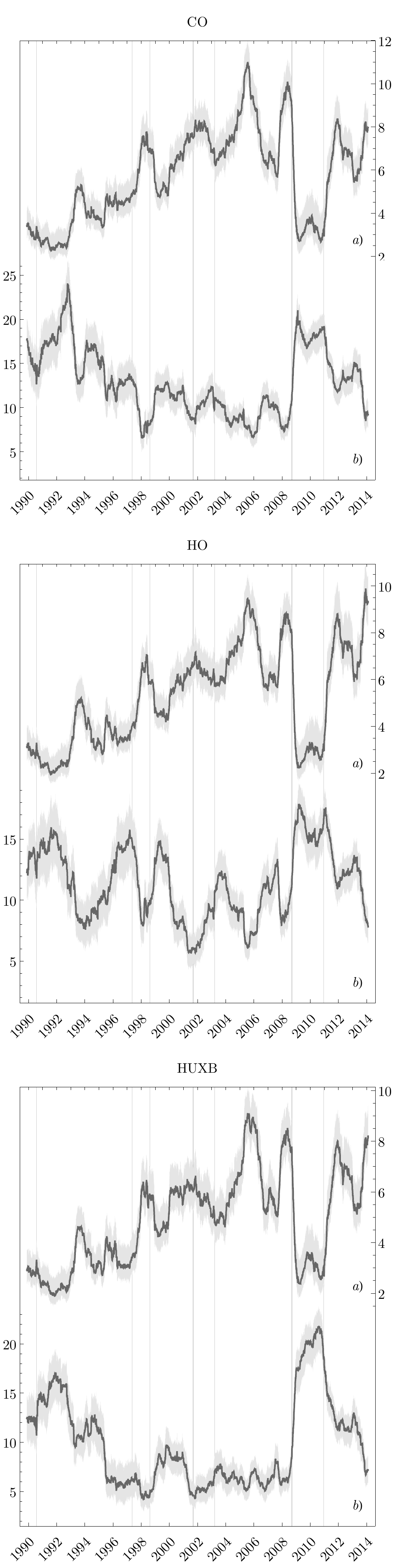}
        \caption{\textsc{To} connectedness}
        \label{fig:toconnectedness}
  \end{subfigure}
    \caption{Directional connectedness of crude oil (CO), heating oil (HO), and gasoline (HUXB). The left figure shows the within connectedness measure, and the right figure shows the frequency connectedness. The top lines denoted as \emph{a)} show the measures on the frequency band of up to a week (one to five days). The bottom lines denoted as \emph{b)} show the measure on the frequency band from a week to two years (six and more days to 500 days). The shaded 10\% confidence bands are based on parametric bootstrap.}
  \label{fig:directionalconnectedness}
\end{figure}

In the Figure~\ref{fig:directionalconnectedness}, we present the disaggregated directional effects of information shocks \textsc{to} and \textsc{from} other elements of the system. 

The most interesting are the relevant figures that investigate the contribution of shocks \textsc{from} other elements to crude oil and \textsc{to} the derivatives from crude oil, as they show how demand-side shocks influence the supply side and how the supply side influences the demand side, respectively.

Starting with the shocks \textsc{from} the products of crude oil to crude oil, depicted in the upper left part of Figure~\ref{fig:fromconnectedness}, it is apparent that within the frequency band of up to one week, the crude oil volatility is increasingly influenced by the shocks from the demand side (from the volatility of the other two assets). Numerically, this results in an increase in within connectedness from 8\% to 20\% over the course of 20 years. This picture is consistent with possible financialization supporting \citep{singleton2013investor,masters2008accidental}, as big financial institutions are much more sensitive to movements in volatility and also willing to rebalance their positions more often. Long-term connectedness, however, remains relatively stable until the economic crisis of 2008. As gasoline and heating oil are products of oil and signify demand after products that are produced in the same production process over many years, it would be surprising if the long-term impact of shocks should be significantly changing under conditions of a stable economy. With the advent of crisis, however, the demand shocks become an important signal about the state of economy for the supply side, hence the increased within connectedness in both the short and long runs.

These findings are complemented by the figure describing how the shock \textsc{to} crude oil influences the other two assets in total, depicted as the upper right picture of the Figure~\ref{fig:toconnectedness}; there is only a slight increase in connectedness over the pre-2008 period, with a high increase during the crisis in the short run. This means that any information shocks into crude oil get transmitted to the products of crude oil very quickly. On the contrary, in the long run, the connectedness trends downward, most probably signifying that the economies are nowadays better able to offset any long-run effects of shocks to oil by switching to other fuels.

The rest of the plots support the previous findings. Shocks \textsc{from} the other two assets to heating oil and gasoline in the short run move around without a significant trend over the pre-crisis period. This means that neither the crude oil nor the complementary products affect the given asset in profoundly different ways over time. The overall changes are hence mostly caused by the amount of variance that is concentrated on various parts of the spectra. Shocks \textsc{to} the crude oil and complementary asset are significantly tending upwards in the short run, reflecting the increase in the influence of product volatility for the volatility of the crude oil.

\subsection{Supply side vs. demand side shocks}
\label{sub:Supply vs. demand shocks and substitution effects}

\begin{figure}[pt]
  \center
  \includegraphics[scale= 0.47]{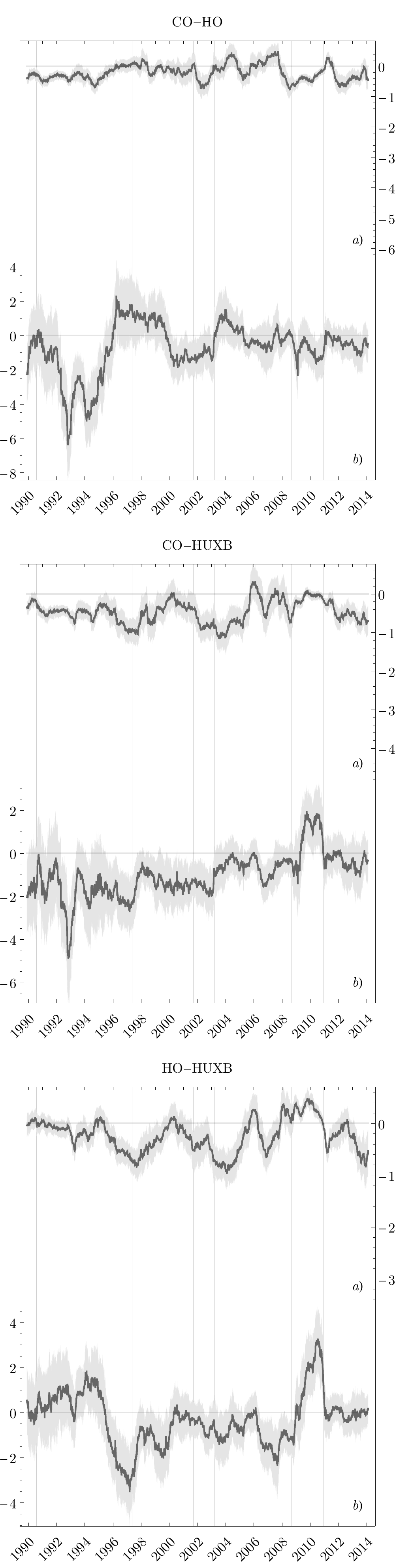}
  \includegraphics[scale= 0.47]{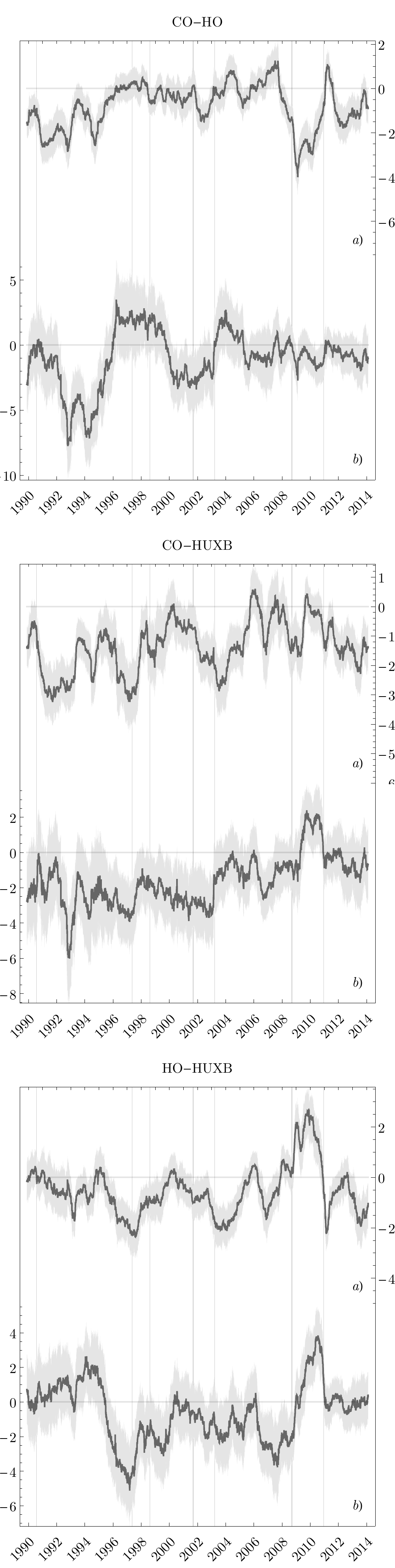}
  \caption{Pairwise within connectedness of crude oil (CO), heating oil (HO), and gasoline (HUXB). The left column shows the within connectedness measure, and the right column shows the frequency connectedness. The top lines denoted as \emph{a)} show the measures on the frequency band of up to a week (one to five days). The bottom lines denoted as \emph{b)} show the measure on the frequency band from a week to two years (six and more days to 500 days). The shaded 10\% confidence bands are based on parametric bootstrap.}
  \label{fig:pairwise}
\end{figure}

As suggested earlier, the system is useful for showing the cyclical properties of supply and demand shocks. In particular, using the pairwise connectedness, we can concentrate on pairs of variables and see which of the two shocks is more important at a given time. All the \textsc{pairwise} connectedness is depicted in Figure~\ref{fig:pairwise}.

We start with the pair of crude oil and heating oil. If the pairwise connectedness from this pair is negative, it means that the shock to crude oil influences heating oil more than shock to heating oil influences crude oil. In the short term, while slightly significant, the within connectedness is very close to zero, meaning that both shocks to heating oil (demand side) and to crude oil (supply side) have similar effects on each other. However, the within connections in the long run are much more pronounced. During the period of 1992-1996, the shocks to crude oil (the supply side) influence the demand side much more than vice versa. The same is true but less pronounced during the period of 2000-2004.

Turning to the crude oil and gasoline pair, in the beginning of the sample, the supply side shocks are more important for the volatility of gasoline than vice versa. However, during the period of 2006-2009, the gasoline shock becomes more important in the short run, and this is followed in the period 2009-2011 by dominance of the gasoline shocks in the pair in the long run.

The last pair including heating oil and gasoline reveals yet another dependence pattern. From the beginning of the sample until year 1996, the relative importance of the shocks is mostly non-distinguishable from zero. In the period 1996-2000, both in the long run and in the short run, the heating oil dominates in terms of the importance of the shocks in the pair. Another important peak is in the long run in the period 2009-2011.

The absolute connectedness of the system is in most cases only an amplified version of the within connectedness, as a brief glance at the right column warrants. Hence, in relative terms, the structure of the within connectedness is the driving force behind the time dynamics.

\subsection{Short term and long term importance}
\label{sub:Short term and long term importance}

\begin{figure}
  \center
  \includegraphics[scale= 0.35]{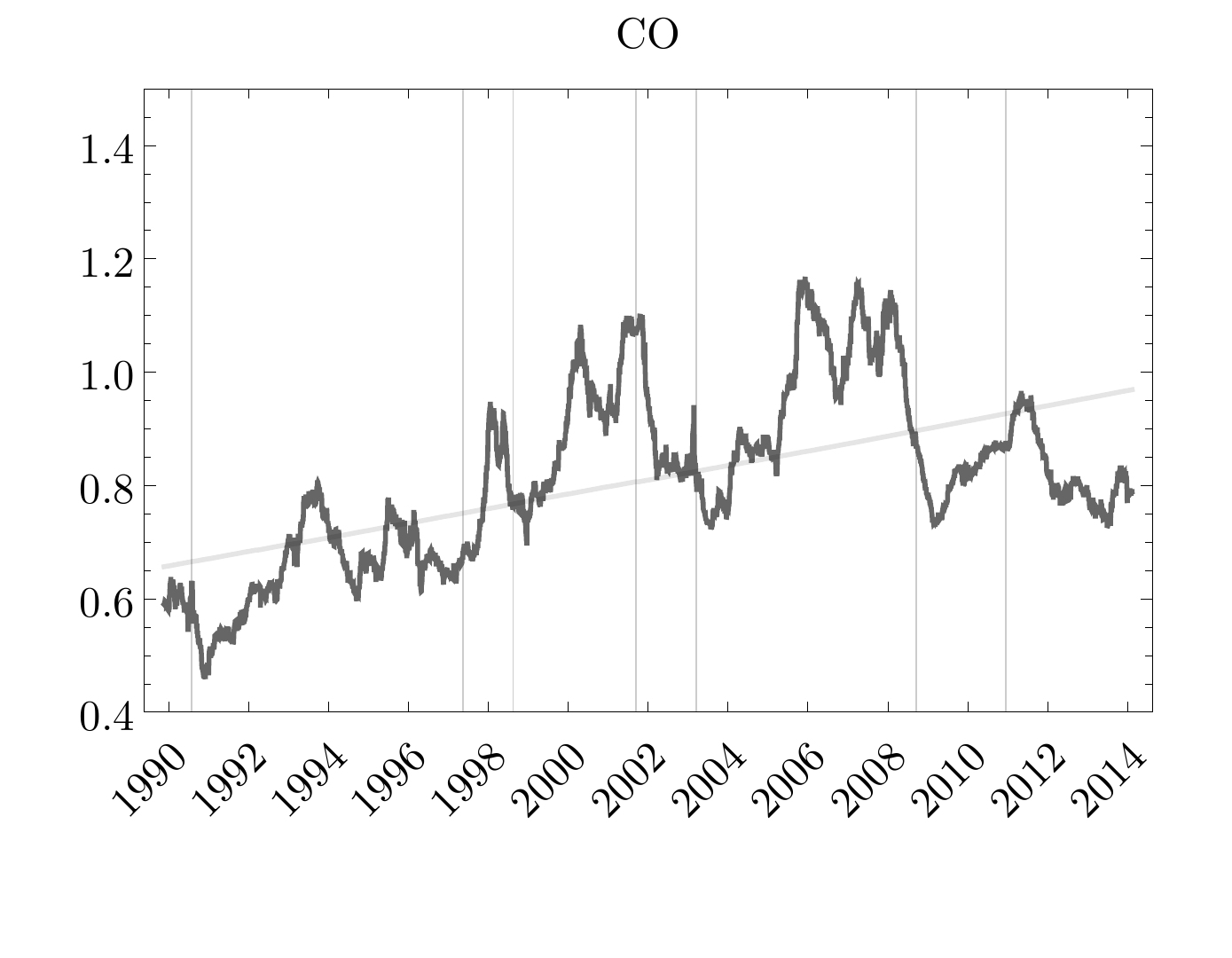}
  \includegraphics[scale= 0.35]{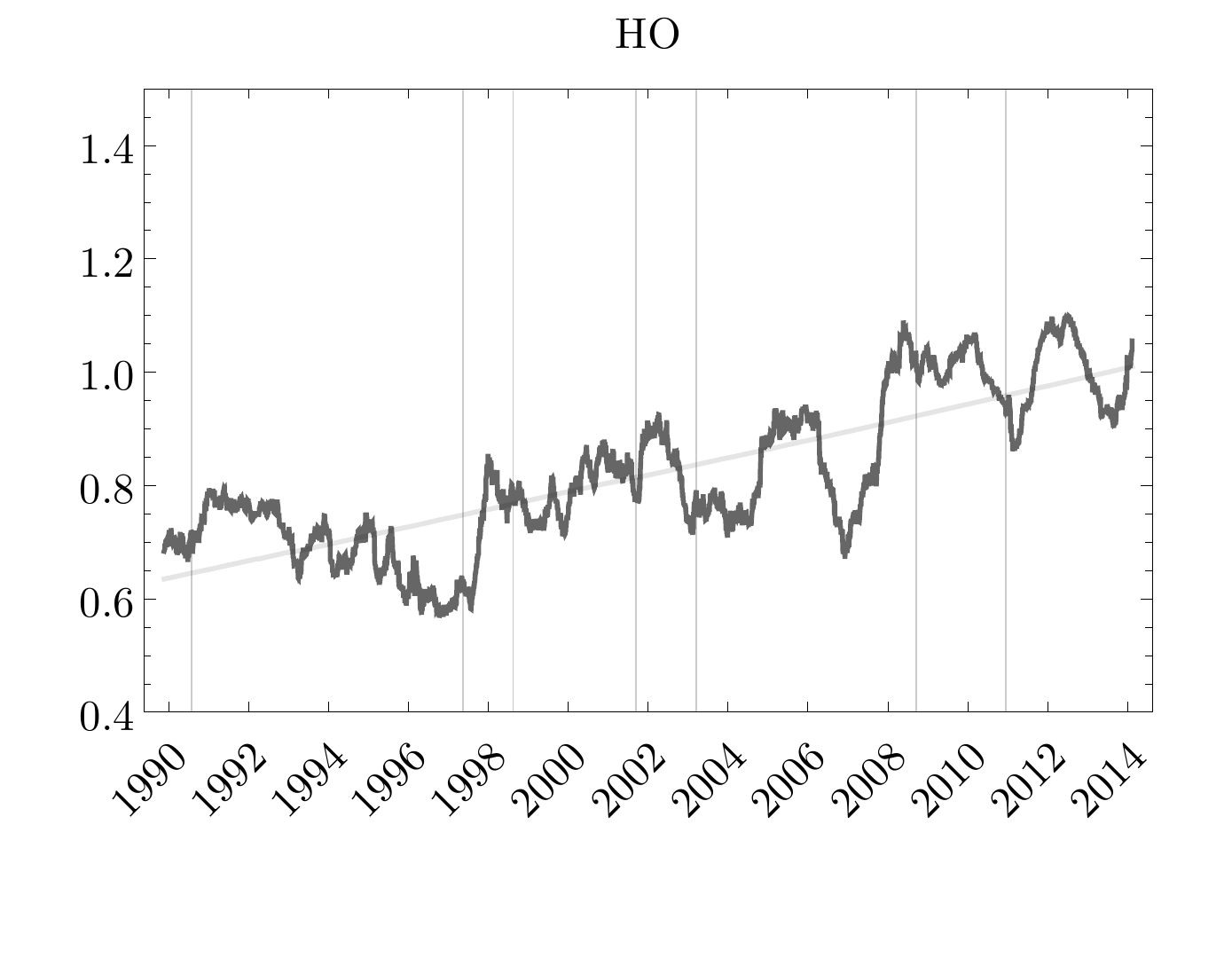}
  \includegraphics[scale= 0.35]{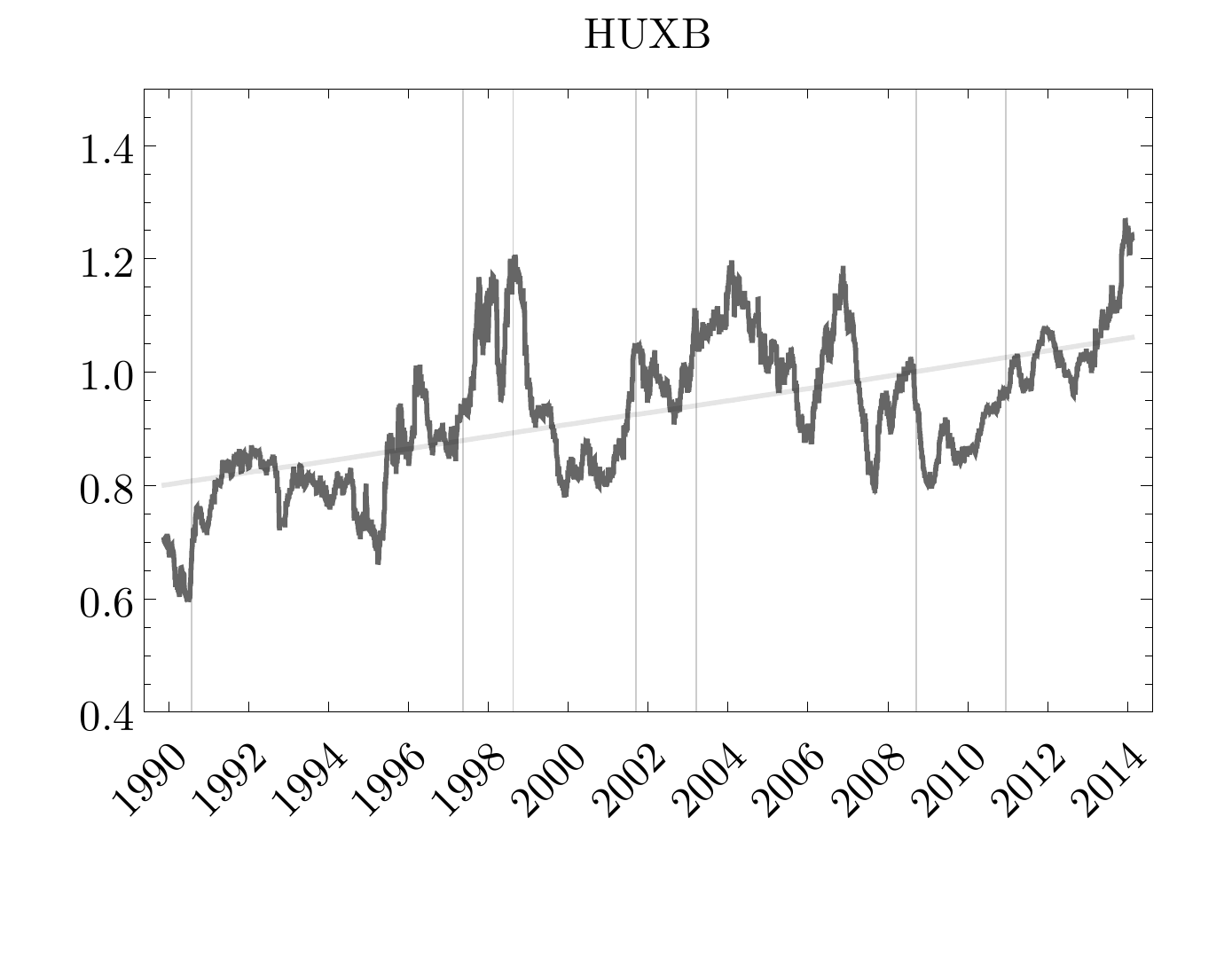}
  \\
  \includegraphics[scale= 0.35]{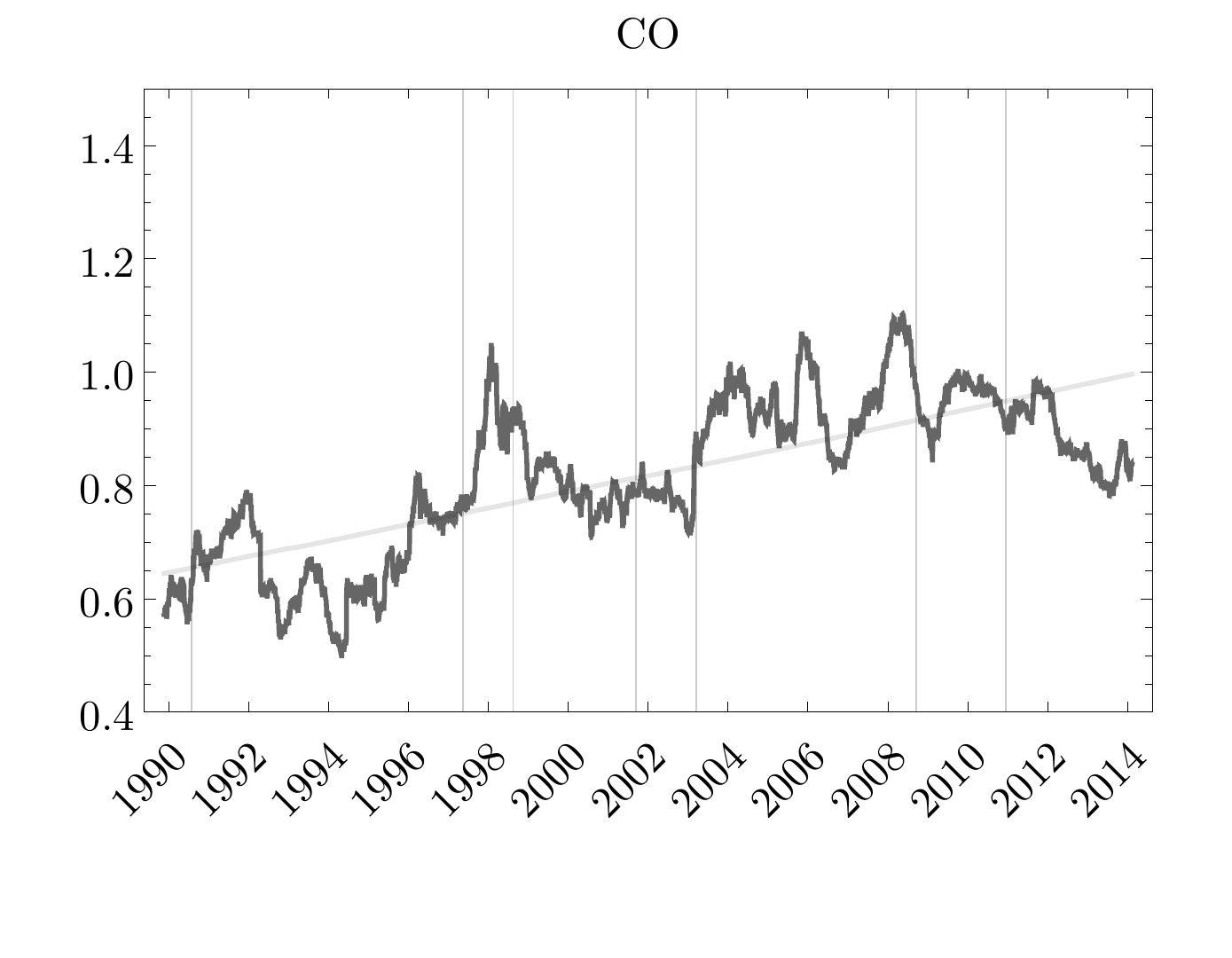}
  \includegraphics[scale= 0.35]{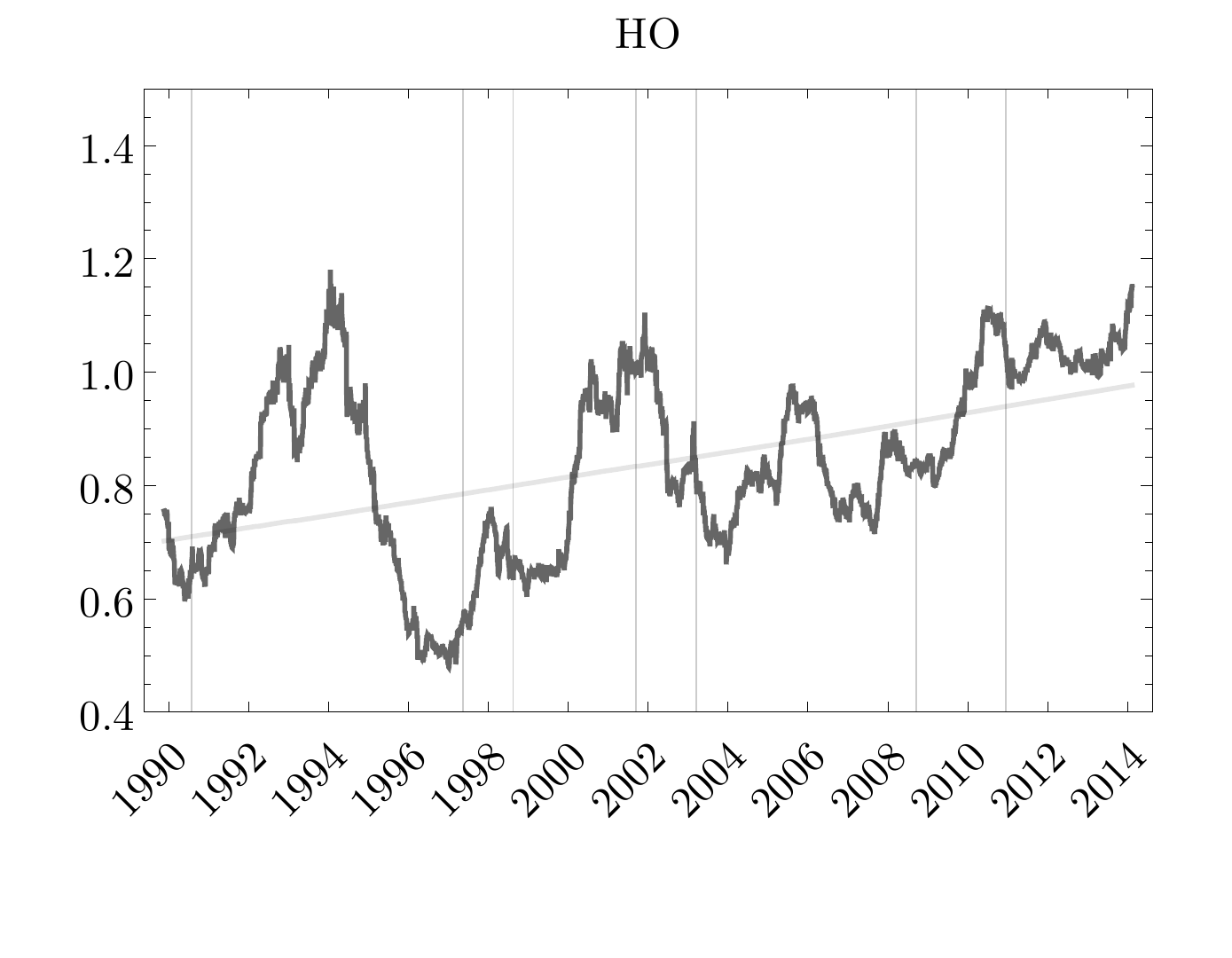}
  \includegraphics[scale= 0.35]{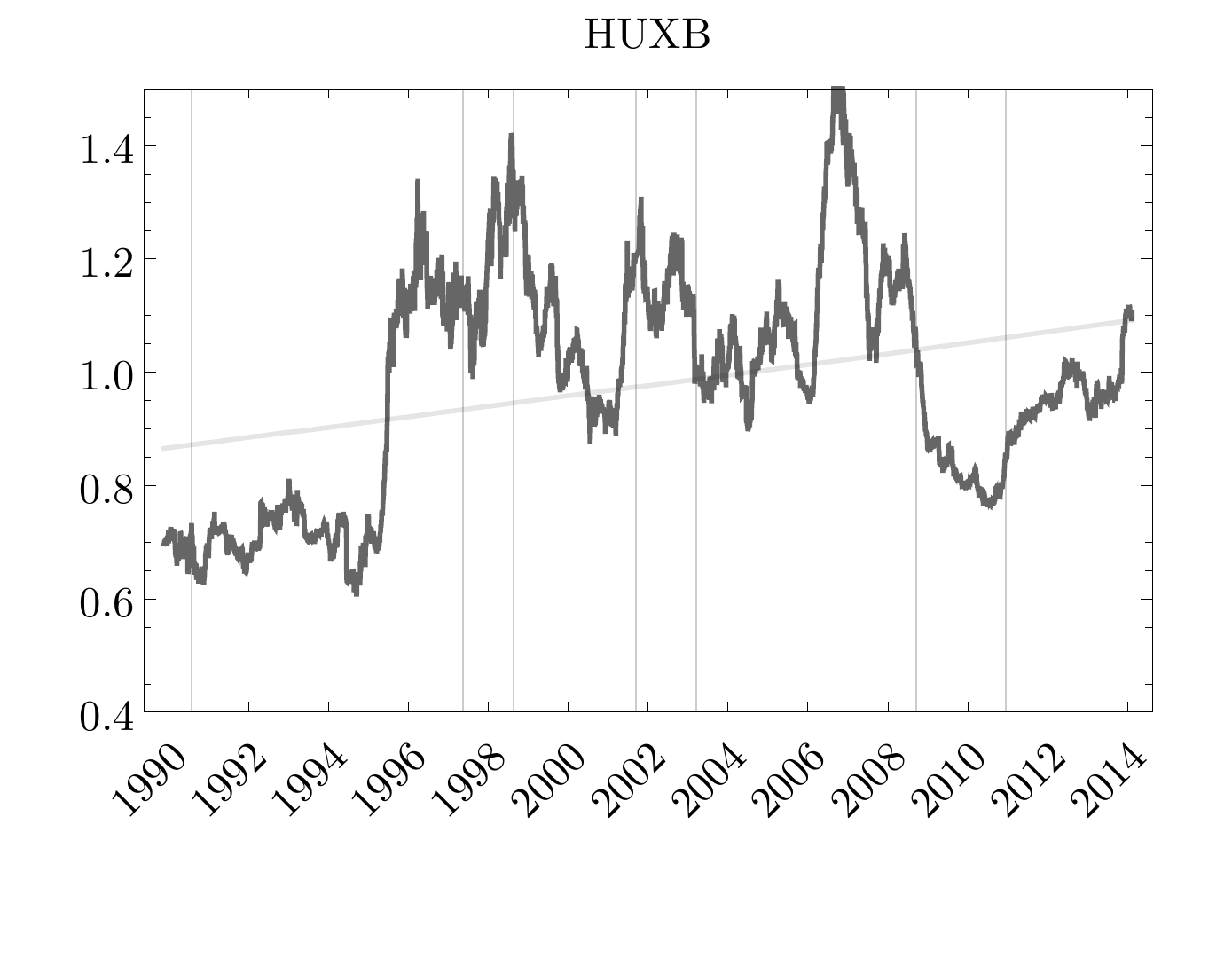}
\caption{Ratio of within connectedness with superimposed linear fit. The first row contains the \textsc{to} connectedness, and the second line contains the \textsc{from} connectedness.}
  \label{fig:ratios}
\end{figure}

Interestingly, in all the observed measures, the importance of the short-term component, \emph{i.e.,} movements up to a week, considerably increase over time both in the within connectedness and in the frequency connectedness. We demonstrate this in Figure~\ref{fig:ratios} that shows the ratios for within \textsc{from} and within \textsc{to} connectedness for all three commodities with superimposed linear regression over the whole sample.\footnote{The ratios in the non-within measure suggest the same qualitative interpretation. The reader can refer to Figure~\ref{fig:ratiosabs}.}

Hence, a singular shock to either one of the oil-commodities is more prone to causing movements that clear within one week’s time. With the weights following a similar pattern, this is good news for long-term investors, as their exposure to systemic risk over the long term has decreased. We may only hypothesize over the sources of the profound change. The market may have become more efficient in realizing and quickly clearing the price changes, which would anecdotally point to higher involvement of financial institutions. Should the world economy become less dependent on oil in the long run or experience less uncertainty about the sufficiency of future supplies of oil, shocks to volatilities will have lower long-term effects than they would have otherwise.

Despite concentrating on the within frequency connectedness, as it pertains more to the fundamental understanding of the risk transmission regardless of the frequency properties within the window, our conclusions would only be stronger. Because the spectral weights of the system increase pre-crisis, we would obtain even steeper estimates of the growth of relevance pre-crisis. During the period 2008-2014, the results would be hazier because of the extreme weights on the long-term movements.

\section{Conclusions}

The oil commodity markets are currently one of the most important commodity markets, as they hugely influence economies in terms of determining a big share of the prices of transport and energy. In this paper, we illustrate how the energy volatility markets have changed their spectral properties over the last 25 years. Why do we care about cyclical properties of volatility spillovers, and what are the implications for investors, regulators, and facility operators? As volatility is directly translated to risk, substantial changes in volatility and its spillovers across oil-related products are able to negatively impact risk-averse investors. Hence, knowledge about the volatility connectedness at different frequencies has important implications for investors and financial institutions in terms of portfolio construction and risk management at various investment horizons. Additionally, frequency dynamics may be important for accurate asset pricing models and hedging strategies \citep{Dew_Becker_2016}. Because volatility is directly tradable using swaps and futures, it is of direct interest for investors and practitioners to be able to reduce risks with help of diversification. Furthermore, connectedness of volatility is closely related to market co-movements, and this phenomenon becomes pronounced during periods of high uncertainty when an unusually sharp increase in market volatility spills across other markets. Analyzing and measuring the connectedness of volatility due to shocks with heterogeneous frequency responses can provide an ``early warning system'' for crises and map the development of existing crises \citep{diebold2012better}. Proper knowledge of volatility transmission mechanisms then becomes a segment of information that is useful for regulators, operators, and policy makers that may lead to the introduction of regulatory and institutional rules to reduce the cross-market impact of excessive price movements.

In terms of material results, we document increasing importance of the effects of shocks up to one week in overall connectedness, both in the within frequency band and in absolute terms. This increase correlates with the financialization of the commodity markets \citep{tang2012index,buyukcsahin2014speculators}. An increased participation of financial institutions in the commodity markets should cause faster reactions to price shocks because of the exploitation of possible arbitraging opportunities arising from deviations of prices that arise because of the shocks. Such behavior would increase short-term volatility and hence short-term connectedness of the markets.

The long-run effects (one week to two years) of shocks are slowly losing importance over the last two decades, with the exception of the crisis period in 2008 and several subsequent years. Understandably, during the crisis, uncertainty emerges, and any information is processed more carefully. In absolute terms, the connectedness in the short run decreased in the crisis primarily because of long-run changes in the levels of volatility. This finding has an important bearing for systemic risk in cases where oil and its products are involved, as it is more and more important to model the high-frequency aspects of the volatility.

Moreover, we document this trend of growing importance of movements of up to one week across all the directional measures. Hence, this finding not only applies to the system connectedness as a whole but also to how individual assets process risk from the demand or supply side of the market. The change to shorter-run connectedness is proved to be profound and present in all aspects of the system. Finally, we demonstrate that the supply-side shocks dominate in terms of how strong the elicited responses are only in several cases in the dataset, and it is only on rare occasions that demand-side shocks dominate the supply-side shocks. Such occurrence only happened during the period 2006-2008 in the short term, confirming the findings of \citet{kilian2014role} that demand signals were driving the market during that period.

While the methodology is flexible enough, it inherits the limits of the classical vector autoregression framework; hence, one should carefully treat the estimation procedure before blindly interpreting the resulting connectedness index. Nevertheless, an applied econometrician aware of the classical time series procedures may enjoy new developments waiting for discovery, as it is tempting to look at the important problems discussed in the previous paragraphs of the conclusion with the lens of the frequency tools provided in this paper. In this respect, our work opens many interesting avenues to be explored.

{\footnotesize{
\setlength{\bibsep}{3pt}
\bibliographystyle{chicago}
\bibliography{BIBLIOGRAPHY}
}}
\appendix

\begin{figure}
  \center
  \includegraphics[scale= 0.5]{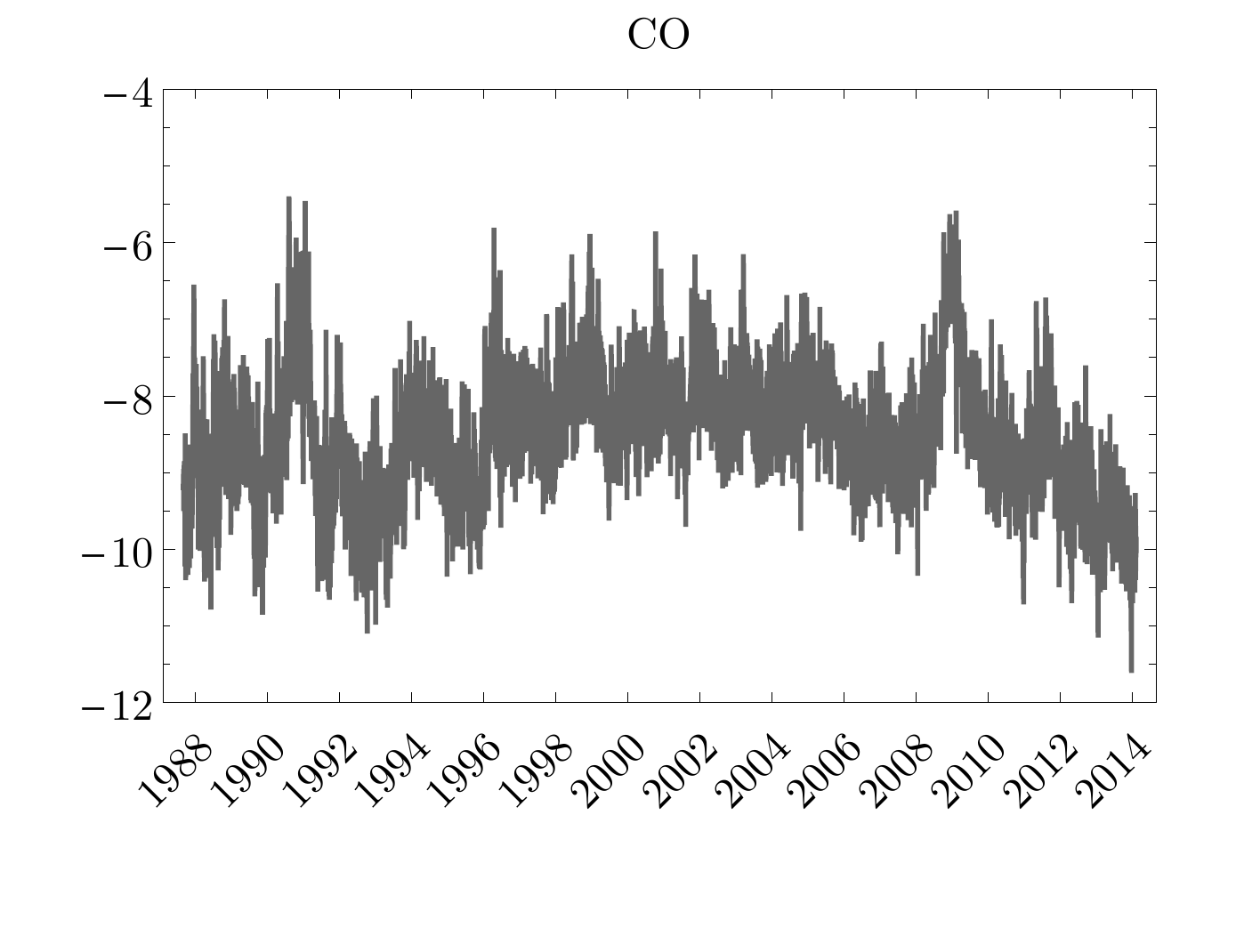}
  \\
  \includegraphics[scale= 0.5]{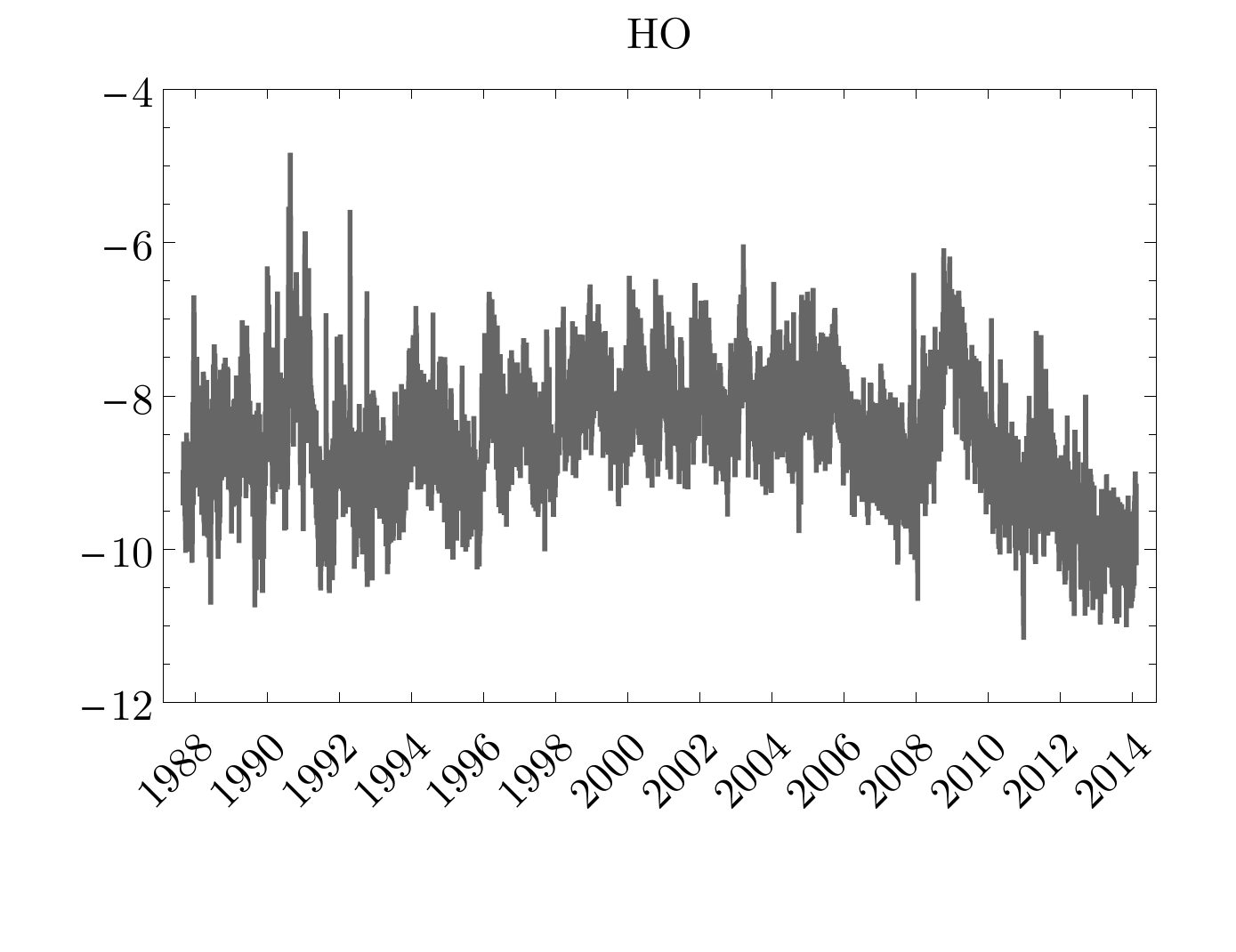}
  \\
  \includegraphics[scale= 0.5]{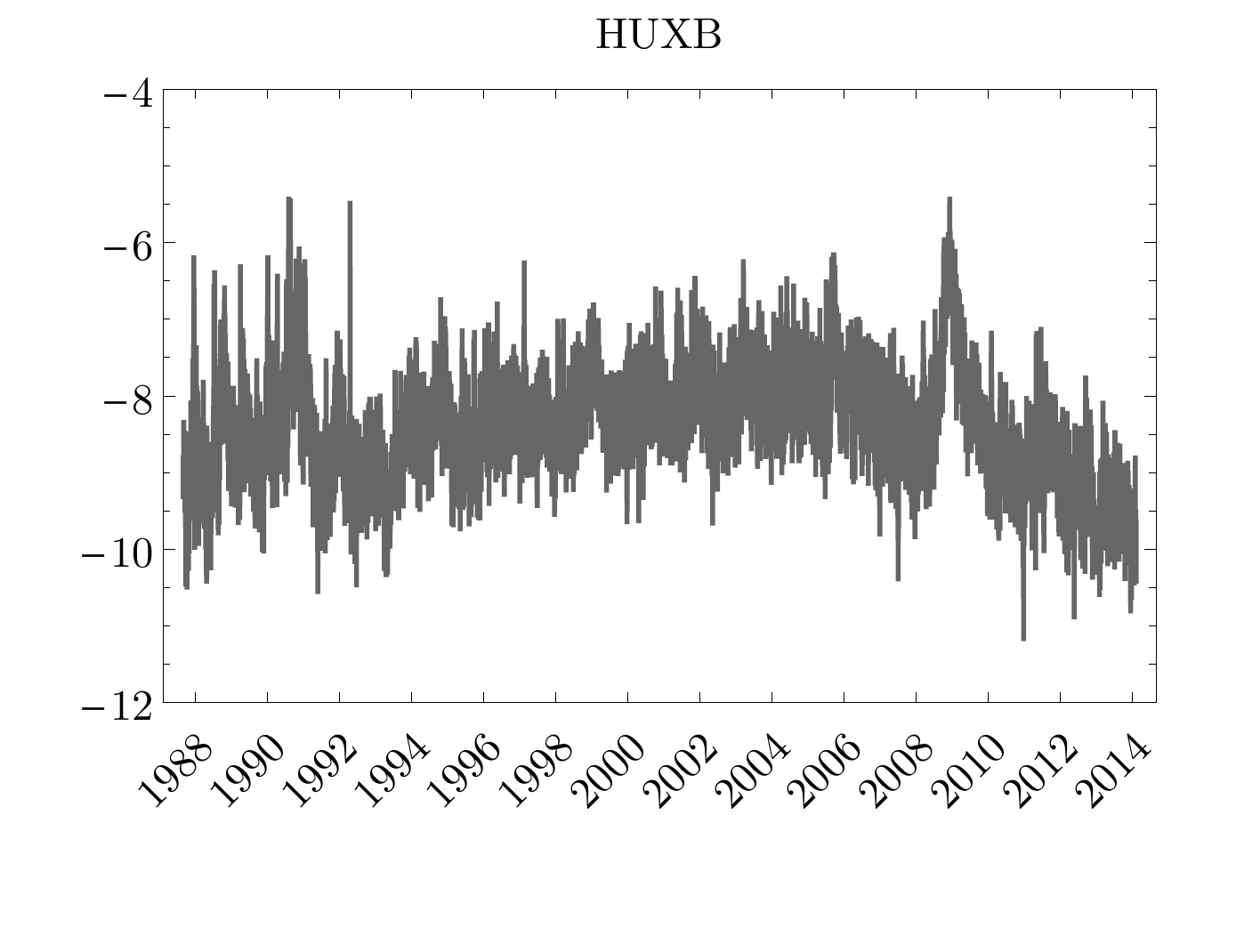}
\caption{Logarithmic bi-power variation.}

  \label{fig:evolutions}
\end{figure}

\begin{figure}
  \center
  \includegraphics[scale= 0.35]{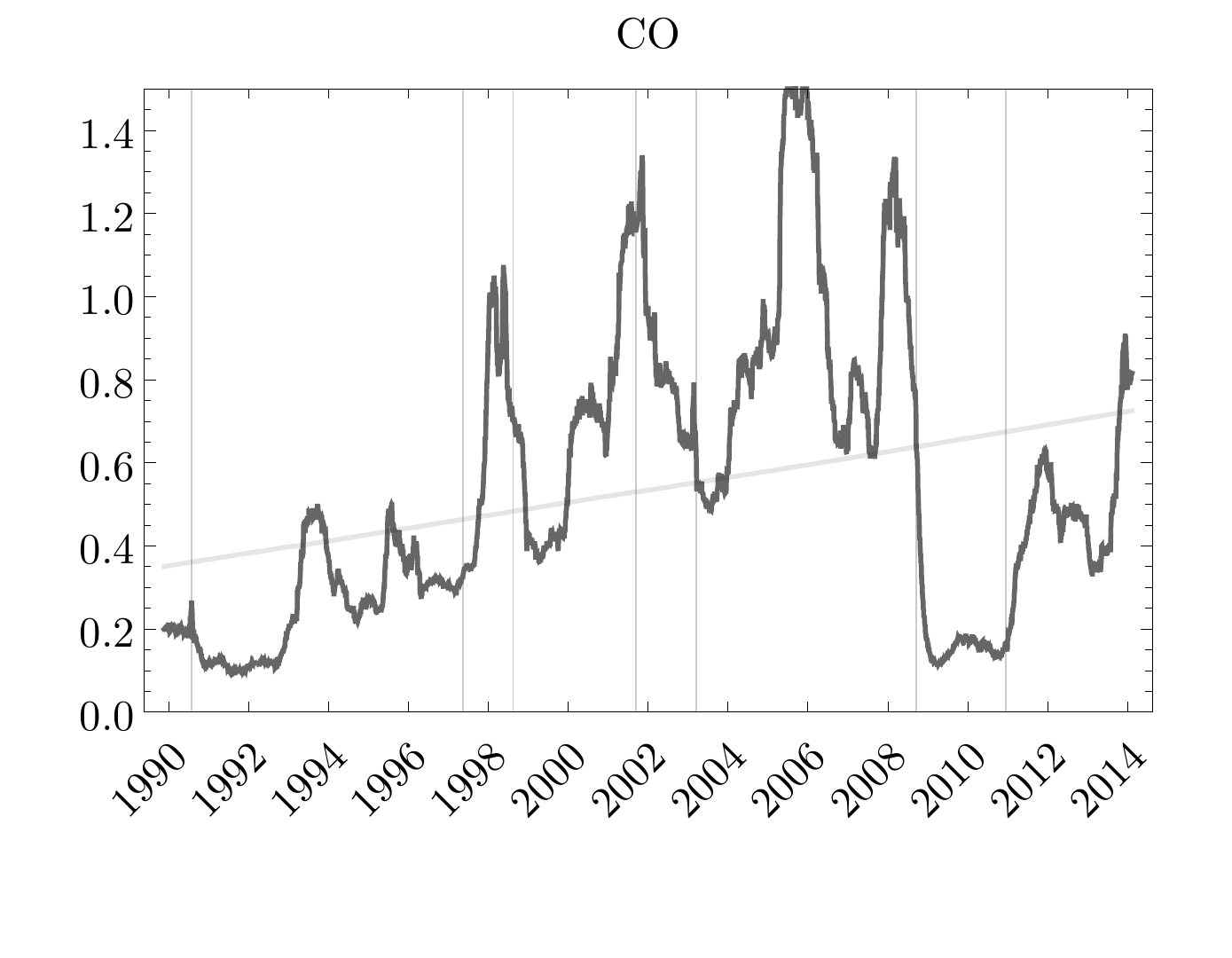}
  \includegraphics[scale= 0.35]{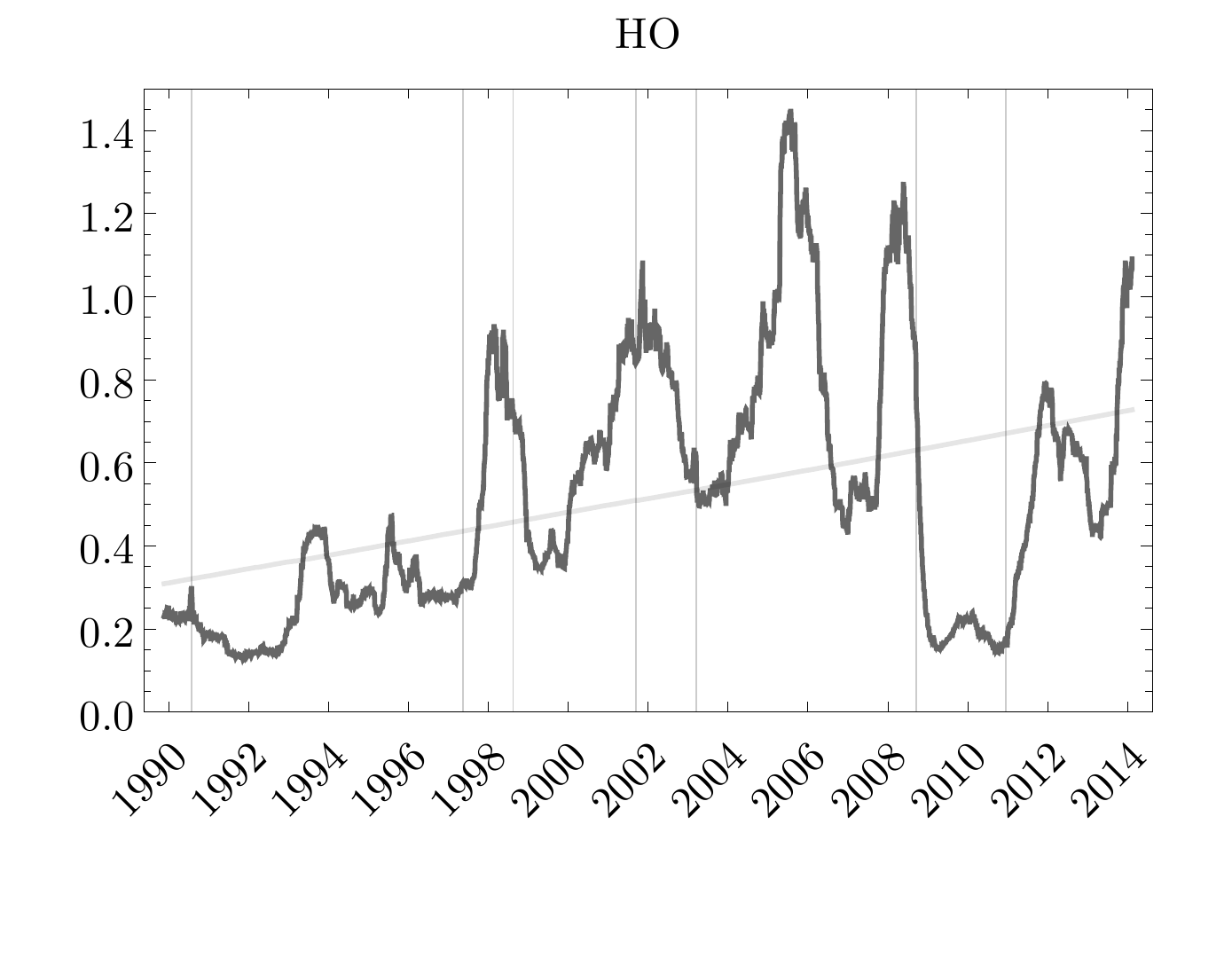}
  \includegraphics[scale= 0.35]{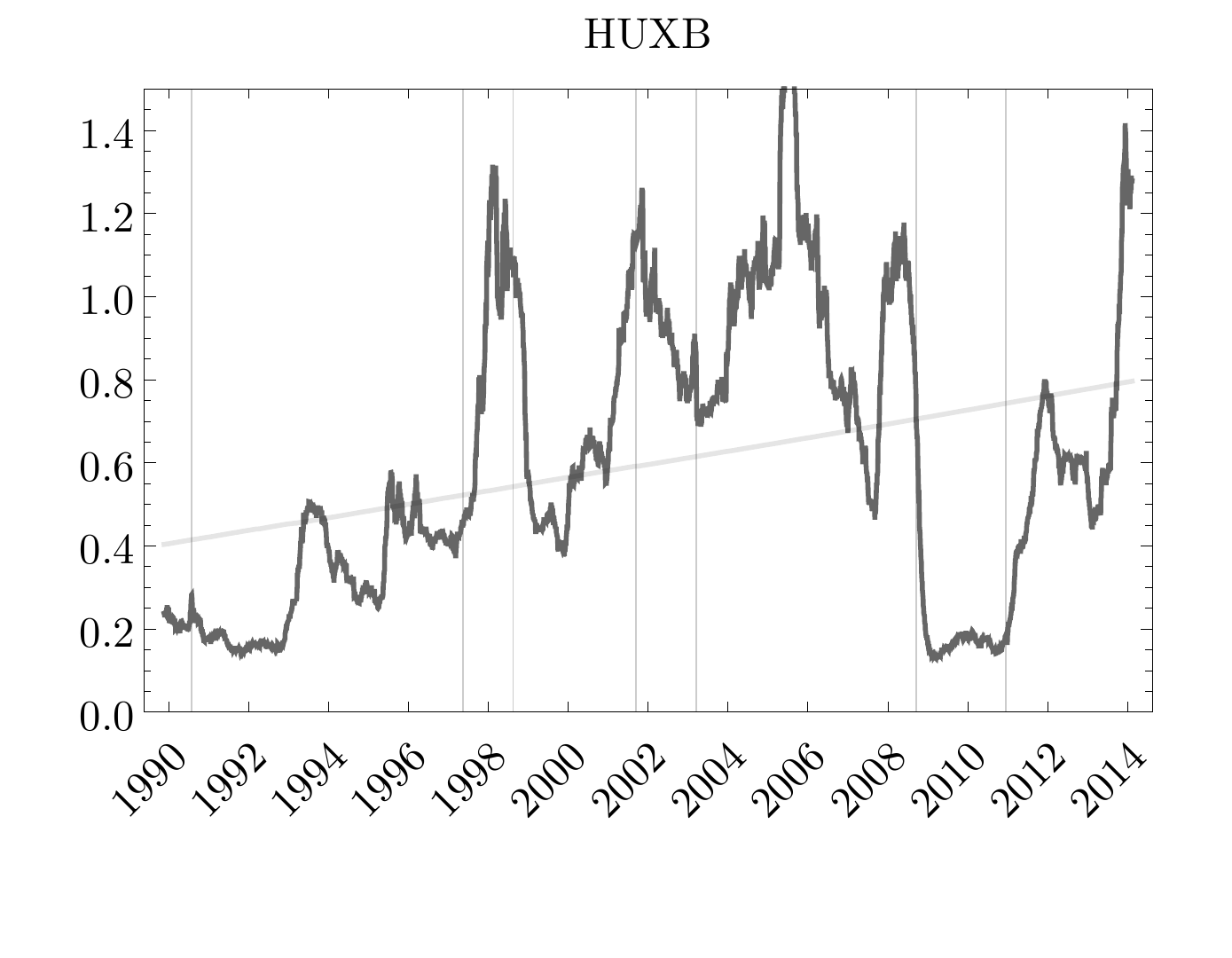}
  \\
  \includegraphics[scale= 0.35]{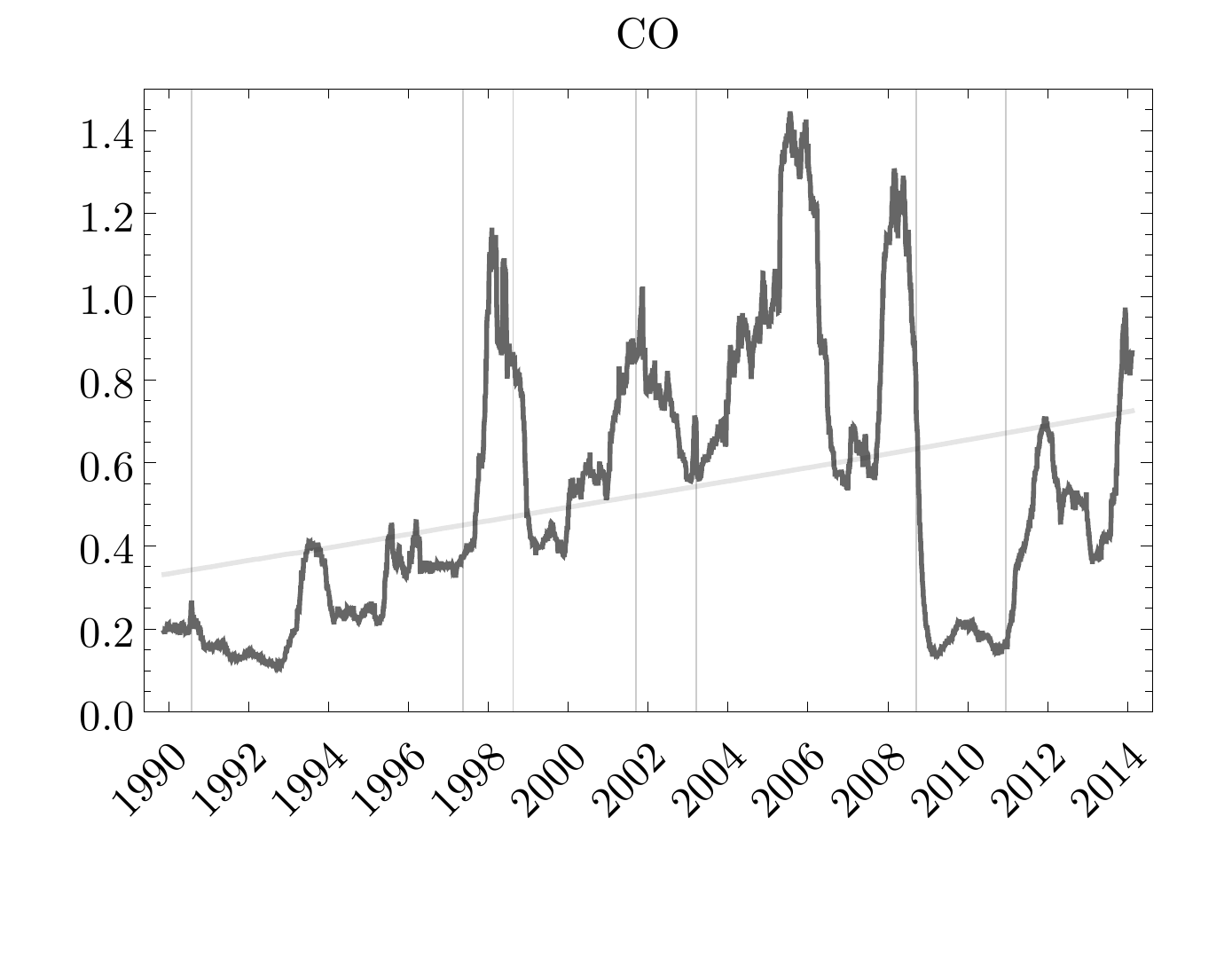}
  \includegraphics[scale= 0.35]{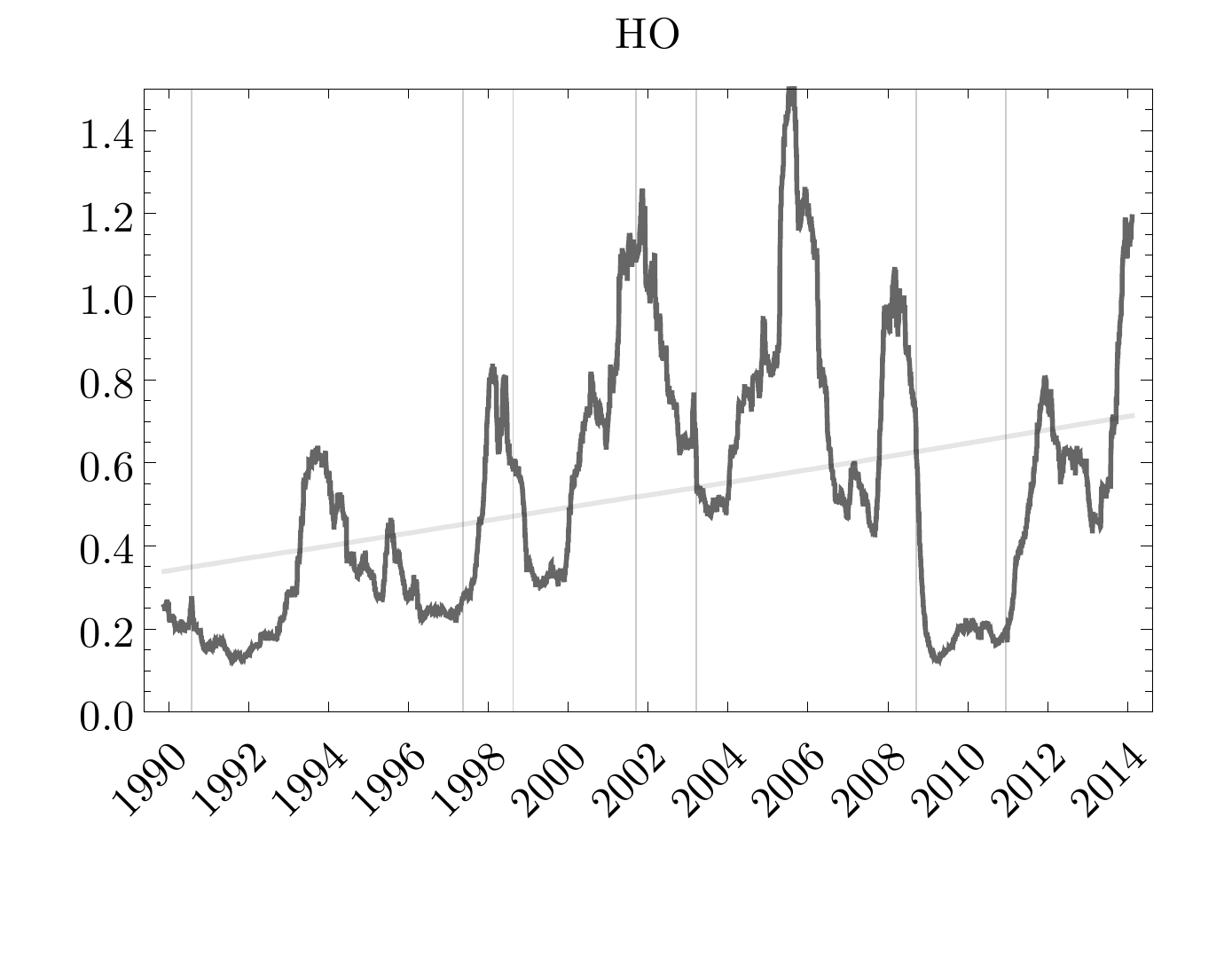}
  \includegraphics[scale= 0.35]{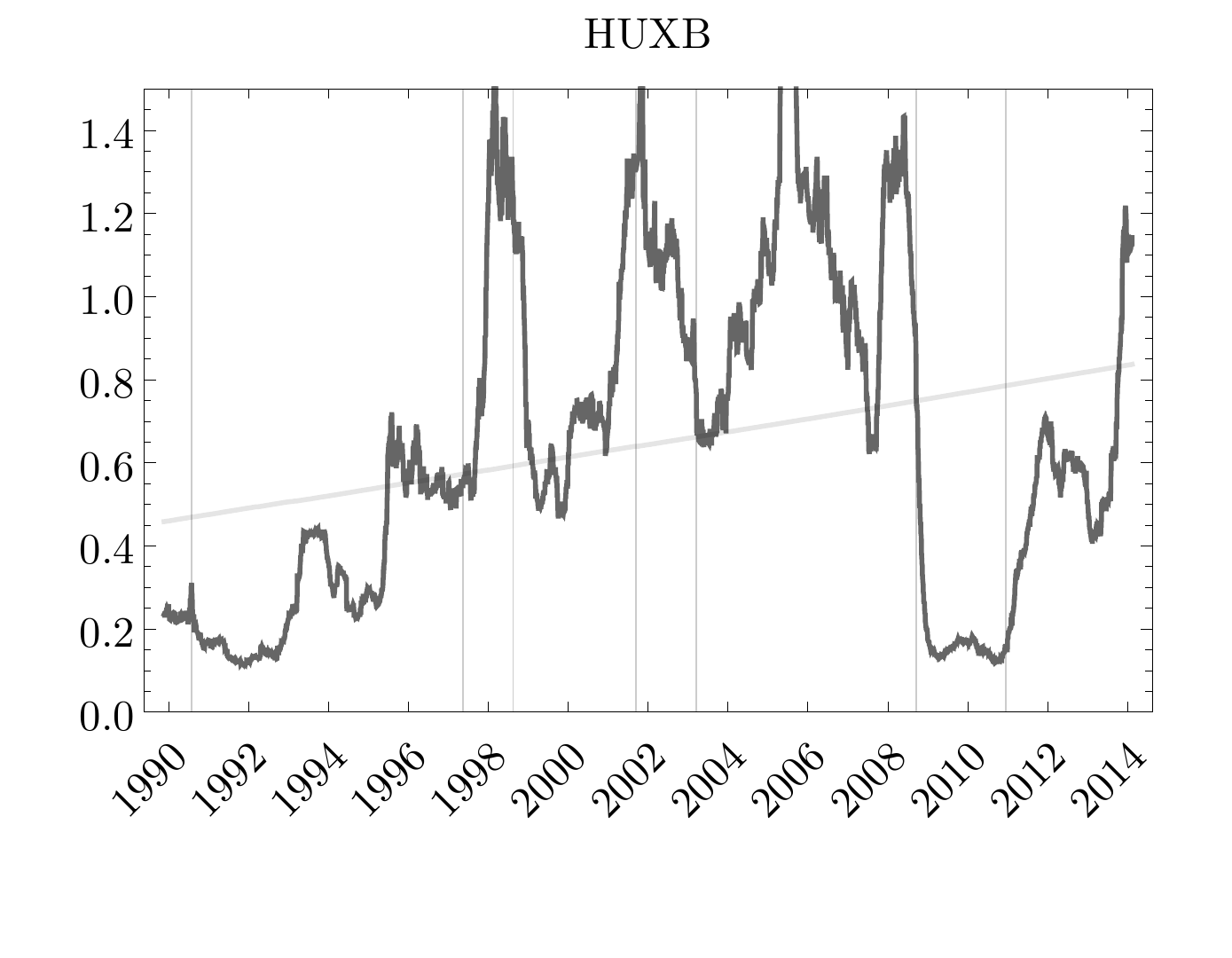}
\caption{Ratio of absolute connectedness to superimposed linear fit. The first row contains the \textsc{to} connectedness, and the second row contains the \textsc{from} connectedness.}
  \label{fig:ratiosabs}
\end{figure}

\begin{table}

\caption{Descriptive statistics of the BPV.}
\centering
\begin{tabular}[t]{lrrr}
\toprule
  & CO & HO & HUXB\\
\midrule
Mean & 0.015 & 0.015 & 0.016\\
Median & 0.014 & 0.014 & 0.015\\
$\sigma$ & 0.007 & 0.006 & 0.007\\
Skewness & 1.799 & 1.430 & 1.725\\
Kurtosis & 5.812 & 5.950 & 5.992\\
\bottomrule
\end{tabular}
\label{tab:desc_stats}
\end{table}

\end{document}